\documentclass[journal,comsoc,onecolumn]{IEEEtran}
\usepackage[utf8]{inputenc}
\usepackage{cite}
\begin{document}

\title{On the feasibility and usefulness of applying the `Schr\"odinger c.q. Liouville-von Neumann equation' to quantum measurement}

\date{12,07,2023}

\author{
    \IEEEauthorblockN{Willem M. de Muynck\\TU/e, Group SMB, FLUX 5.105, Postbus 513, 5600 MB  Eindhoven, The Netherlands \\
      E-mail: W.M.d.Muynck@tue.nl\\}
}
\bibliographystyle{unsrt}
\raggedbottom
\addtolength{\baselineskip}{0.1\baselineskip}

\vspace*{1cm}  \noindent
{\Large\textbf{On the feasibility and usefulness of applying the `Schr\"odinger c.q. Liouville-von Neumann equation' to quantum measurement}}\\
\vspace{1cm}

\noindent{\bf Willem M. de Muynck}\\
TU/e, Group SMB, FLUX 5.105, Postbus 513, 5600 MB  Eindhoven, The Netherlands \\
E-mail: W.M.d.Muynck@tue.nl\\

\begin{abstract}
    The present paper is a sequel to papers \cite{MadM93,dM98,dM2000,NeoCop2004} dealing with recent developments on
the issue of `quantum measurement'.
In this paper `measurement within the domain of application of quantum mechanics' is treated as a \emph{quantum mechanical}
\emph{interaction} of a `(sub)microscopic object $(o)$' and an `equally (sub)microscopic part of the measuring instrument $(a)$
being sensitive to the (sub)microscopic information', that interaction to be described by a Schr\"odinger equation.
The Stern-Gerlach experiment is used as a paradigmatic example. An alternative to the Heisenberg inequality is found,
exhibiting the \emph{independent} contributions of `preparation of the initial state of object $(o)$' \emph{and} `interaction of
object $(o)$ \emph{and} measuring instrument/probe $(a)$'. Applicability of the Liouville-von Neumann equation is stressed.
\end{abstract}


\noindent Statements and Declarations: CC BY \\

\noindent {\bf ACKNOWLEDGMENT}: The author likes to thank Eindhoven University of Technology,
in particular prof. Kees Storm, for their enduring hospitality.
Thanks are also due to Peter Kirschenmann for his insistence on
`within the math keeping visible the difference between object $(o)$ and ancilla $(a)$'.

\section{\bf Introduction}
\label{sec1}

This paper is \emph{not} about `the quantum mechanical measurement problem'.
It is dealing with the problem of `\emph{measurement within the (sub)microscopic physical domain
as far as described by quantum mechanics}'.\footnote{Hence, hidden variables theories are left
undiscussed (see e.g. de Muynck \cite{dM2002}, Chapt.~10 for this latter issue).}
Nowadays it is appropriate to look somewhat differently
on `measurement within the (sub)microscopic physical domain' than did Bohr in his revolutionary `Como lecture \cite{Bohr1928}',
or von Neumann in his, for that time excellent, `Mathematical foundations of quantum mechanics \cite{vN32}'.

Rather than sticking to Bohr's well-known idea that `measuring instruments should be \emph{macroscopic}', nor to von Neumann's conviction that `quantum measurement' and
`free evolution'\footnote{`free' is used here as `having \emph{no} interaction with any
\emph{measuring instrument}'.} are
``two fundamentally different types of interventions'' (cf.~von Neumann \cite{vN32}, p.~351) --\emph{only `free} evolution' being supposed to be described by the
Schr\"odinger equation--,
we are more and more used to look upon `quantum measurement' as a process having \emph{four} phases:\\
1) a `\emph{preparation} (or pre-measurement\footnote{Unfortunately
in de Muynck \cite{dM2002}, \S3.2.1 the term `pre-measurement' was used for the second phase (see also point 2).}
phase' in which an external source (like a cyclotron, a lamp, or the sun)
is \emph{preparing an ensemble} of\footnote{See eqs~(\ref{eq:3.1.5.x}) and (\ref{7a.2.1.2})
for the reason of the addition $(exp)$.} $N_{(exp)}$
``identically'' prepared physical objects $(o)$ (described by the density operator $\rho_{(o)}(t=0)$)
to be submitted to a measurement process'\footnote{I here
take for granted that in actual practice these objects are \emph{not} prepared simultaneously at $t=0$, but \emph{consecutively}.};\\
2) a `\emph{premeasurement} phase'\footnote{Note the difference between `pre-measurement' and `premeasurement', the latter nomenclature
being used by Peres \cite{Peres1980}.}
\emph{quantum mechanically} described by a Schr\"odinger equation in which a `\emph{(sub)microscopic} object $(o)$' is interacting with
an equally `\emph{(sub)microscopic} part of a \emph{measuring instrument} (to be referred to as a \emph{probe} or \emph{ancilla} $(a)$)
being sensitive to the \emph{(sub)microscopic} information' (e.g. Helstrom \cite{Hel76}, p.~74));\\
3) an `\emph{amplification} phase' in which a `\emph{(sub)microscopic} effect of the previous phase' is amplified
to a `\emph{humanly observable} event' usually relying on (semi-)classical theory
(as is the case in the Stern-Gerlach measurement to be dealt with in \S\ref{subsubsec2.1.1});\\
4) a `\emph{registration} phase' in which `a pointer position of a \emph{macroscopic} part of
the measuring instrument' is recorded.

In the present paper it is assumed that \emph{within the quantum mechanical domain} `measurement on a
(sub)microscopic physical object' is described by the Liouville-von Neumann equation. However, in disagreement with most (if not all)
textbooks of quantum mechanics the `interaction of (sub)microscopic object and probe/ancilla' is \emph{not} ignored here.
On the contrary, that \emph{interaction} is thought to be a \emph{crucial} element of `measurement within the domain of quantum mechanics'.
Assuming (sub)microscopic physical object $(o)$ and probe $(a)$ to be \emph{un}correlated at $t=0$, we then have a
Liouville-von Neumann equation according to
\begin{equation}
\frac{d\rho_{(oa)}(t)}{dt} = - \frac{i}{\hbar} [H_{(oa)},\rho_{(oa)}(t)]_-,\; t\geq 0,\;\rho_{(oa)}(t=0)=
\rho_{(o)}(t=0)\rho_{(a)}(t=0),
\label{eq:2.8.10}
\end{equation}
$H_{(oa)}$ the Hamiltonian of the `\emph{interacting} system of (sub)microscopic object $(o)$ and probe $(a)$,
\emph{statistically}\footnote{In this paper quantum mechanics is accepted as a \emph{statistical} theory (see also \S\ref{subsec3.3}).}
described by the density operator $\rho_{(oa)}(t)$'.
This equation\footnote{Its `restriction to the (sub)microscopic object ($o$)'
is occasionally referred to as the Liouville-von Neumann equation. In order not to burden this paper with --for the present purpose--
irrelevant nomenclature, I shall refer to eq.~(\ref{eq:2.8.10}) --\emph{also} if applied to $\rho_{(oa)}(t)$--
as the `Schr\"odinger equation'.} will be the main subject of the present investigation, the final solution of this equation in its general form being given according to
\begin{equation}
\rho_{(oa)}(t=T) = U_{(oa)}(T) \rho_{(o)}(t=0)\rho_{(a)}(t=0) U_{(oa)}(T)^{\dagger},\; U_{(oa)}(T) =
e^{-{\frac{i}{\hbar}} H_{(oa)}T},
\label{eq:3.1.1}
\end{equation}
$T$ the `\emph{interaction} time of object $(o)$ and ancilla $(a)$'.\footnote{It is assumed that
`measurements within the domain of application of quantum mechanics' are \emph{stochastic} processes (see, for instance,
Davies \cite{Davies1969}, also Nagasawa \cite{Nagasawa2000}, for the appropriateness of the qualification `stochastic')
having (sub)microscopic measurement times symbolized by $T \rightarrow \infty$.}

In the following we will be interested in the \emph{post}-measurement state $\rho_{(a)}(t=T) =Tr_{(o)}(\rho_{(oa)}(t=T))$
of the ancilla $(a)$, depending on the `\emph{measurement interaction} of $(o)$ and $(a)$', and, hence, \emph{also} depending on
the initial state $\rho_{o}(t=0)$ of object $(o)$.
Our point of departure will be the `relative frequencies of the \emph{post}-measurement pointer positions of the \emph{ancilla} $(a)$'
(which are \emph{the only experimental data made available by measurements within the quantum mechanical domain}),
from which we hope to obtain `\emph{knowledge} about the \emph{pre}-measurement density operator $\rho_{o}(t=0)$ of object $(o)$'.
`Object $(oa)$', --consisting of both $(o)$ and $(a)$--,
is considered to be a `\emph{necessary} part of the \emph{domain of application of quantum mechanics}'.

By now it is well-known that `properties of (sub)microscopic objects' need `\emph{positive operator-valued measures
(POVMs)}' for their description (e.g.~\cite{povm}, see also `\S\ref{subsubsec2.1.1} (below)' for a paradigmatic example).
Hence, \emph{rather than} the \emph{standard} probabilities
\begin{equation}
\begin{array}{l}
{\bf p}_{(o)j}=Tr_{(o)}\rho_{(o)}(t=0)E_{(o)j},\; j=1,\ldots, N({\cal H}_{(o)}), \\
E_{(o)j}= |(o)j\rangle\langle(o) j|\; {\rm projection \; operators \; on \; states}\; |(o)j\rangle,
\label{eq:3.1.0}
\end{array}
\end{equation}
$N({\cal H}_{(o)})$ the number of mutually orthogonal states $|(o)j\rangle$ spanning the
`Hilbert space ${\cal H}_{(o)}$ of all possible quantum states $|\psi_{(o)}\rangle$ of object $(o)$
of \emph{standard} quantum mechanics', the set $\{E_{(o)j}, j=1,\ldots, N({\cal H}_{(o)})\}$ defining a
`\emph{projection-valued} measure (\emph{PVM}) (restricted to in the first half of the $20^{th}$ century)',
we have to deal with `probabilities of \emph{final pointer positions} \emph{of an ancilla} $(a)$'
quantum mechanically described by
\begin{equation}
\begin{array}{l}
{\bf p}_{(a)k}(T) := Tr_{(a)} (\rho_{(a)}(t=T)M_{(a)k}), \; \rho_{(a)}(t=T) =Tr_{(o)}(\rho_{(oa)}(t=T)), \\
\{M_{(a)k}, \; k=1,\ldots,N_{(a)}\} \; {\rm satisfying} \; M_{(a)k}\geq O_{(a)},\; \sum_{k=1}^{N_{(a)}} M_{(a)k}=I_{(a)},
\end{array}
\label{eq:3.1.2}
\end{equation}
$\rho_{(oa)}(t=T)$ being given by (\ref{eq:3.1.1}), the set $\{M_{(a)k},\; k=1,\ldots,N_{(a)}\}$
defining a POVM\footnote{Note that physicists (including myself)
often are a bit sloppy about the notion of a `POVM', \emph{not} distinguishing between a
`(non)orthogonal decomposition of the unit operator $I$ (NODI), consisting of $N_{(a)}$ observables', and
a `POVM (being the set of all sums of $k \;(0 < k \leq N_{(a)})$ elements selected from a NODI').},
$T$ the `time at which the measuring instrument's \emph{pointer} reached one of its $N_{(a)}$ (different) \emph{post}-measurement \emph{pointer} positions'.

By changing the `object of interest' from `$(o)$' to `$(oa)$'
the `\emph{generalized} formalism of quantum mechanics represented by (\ref{eq:3.1.2})'
is able to deal with the problems met by the `\emph{standard} formalism' (see e.g. de Muynck \cite{dM2002},
also \S\ref{subsec3.5} of the present paper), probabilities being \emph{restricted} to the
`relative frequencies $Tr_{(o)}(\rho_{(o)}(t=0) M_{(o)k}(T))$ of the \emph{(sub)microscopic object} $(o)$,
$\{M_{(o)k}(T)\}$ a POVM of the \emph{latter} object, to be defined below (cf.~eq. (\ref{eq:3.1.3.2.x})).
In the following `object $(oa)$' --consisting of both $(o)$ and $(a)$--
is considered to be a `\emph{necessary} part of the \emph{domain of application of quantum mechanics}'.
Eq.~(\ref{eq:3.1.1}) is yielding the mathematical means to realize a `paradigm shift' from the `\emph{standard} formalism'
to a `\emph{generalized} formalism dealing with the \emph{interaction of (sub)microscopic object $(o)$ and \emph{ancilla} $(a)$}'.

As is customary in most of the quantum mechanical literature of the second half of the $20^{th}$ century, `probabilities ${\bf p}_{(a)k}(T)$ of
the \emph{final} state of the \emph{ancilla} $(a)$' are \emph{interpreted} as `probabilities ${\bf p}_{(o)k}$ of the \emph{initial} state of
the \emph{(sub)microscopic object} $(o)$'. Thus
\begin{equation}
\begin{array}{l}
{\bf p}_{(a)k}(T)= Tr_{(a)}(\rho_{(a)}(t=T) M_{(a)k})=\\
Tr_{(oa)} (U_{(oa)}(T) \rho_{(o)}(t=0)\rho_{(a)}(t=0) U_{(oa)}(T)^{\dagger}\;M_{(a)k}) \equiv \\
{\bf p}_{(o)k} := Tr_{(o)} (\rho_{(o)}(t=0) M_{(o)k}(T)),
\end{array}
\label{eq:3.1.3.2.x}
\end{equation}
the set $\{M_{(o)k}(T), \; k = 1,\ldots, N_{(o)}\}$ generating a `positive operator-valued measure (POVM),
($N_{(o)}$ the number of `\emph{(observationally distinguishable) final} pointer positions in the ensemble)'.\footnote{The present paper is restricting itself
to `(non-)ideal measurements of \emph{PVM}s' (cf.~(\ref{7a.2.1.2}) and (\ref{7a.2.1.1})), for which
$\{M_{(o)k}(T))\}$ can easily be found (cf.~\cite{dM2002}, Chapt.7,8).}
The requirement that `probabilities ${\bf p}_{(o)k}$ and ${\bf p}_{(a)k}(T)$ as expressed by (\ref{eq:3.1.3.2.x}) are equal'
is important because it enables to obtain `knowledge on the \emph{initial} state $\rho_{(o)}(t=0)$ of \emph{object} $(o)$' by
`\emph{observing} the \emph{final} measurement results $Tr_{(a)}(\rho_{(a)}(t=T) M_{(a)k})$ of the \emph{ancilla}'.
Relying on the `\emph{crucial role played} by the interacting pair of (sub)microscopic object (o) and ancilla $(a)$',
it is enabling to stress the \emph{difference} of the `\emph{present} account of quantum measurement' from
`quantum theory as dealt with in the second half of the $20^{th}$ century (including the author's \cite{dM2002})'.
As a matter of fact, the `\emph{numerical equality} of probabilities ${\bf p}_{(o)k}$ and ${\bf p}_{(a)k}(T)$'
is reducing the so-called `quantum mechanical measurement problem
of \emph{the latter period}' to a `\emph{general problem of measurement within the (sub)microscopic domain}' (see also \S\ref{subsec3.4}).

This reduction is obtained by taking seriously the \emph{necessity} of relying on `\emph{measuring instruments
producing knowledge on object $(o)$ by interacting with it} (as represented by (\ref{eq:3.1.1}))',
and `delivering a \emph{numerical} account of that
knowledge by observing the measuring instrument's \emph{final pointer positions} on a pointer scale (as
represented by (\ref{eq:3.1.2}))' (see also \S\S\ref{subsec3.3} and \ref{subsec3.4}
for a discussion of the `\emph{physical in}equality of
${\bf p}_{(o)k}$ and ${\bf p}_{(a)k}(T)$ as dealt with in (\ref{eq:3.1.3.2.x})', this
`\emph{physical inequality}' \emph{existing together with} the `\emph{numerical equality}
of these probabilities').
This `(\emph{theoretical}) equality' should be compared with `\emph{experimentally obtained}
results ${\bf p}_{(exp)k}$'\footnote{See for the \emph{experimental} meaning of ${\bf p}_{(exp)k}$
the example presented in eq.~(\ref{7a.2.1.4}).} according to
\begin{equation}
\begin{array}{l}
{\bf p}_{(a)k}(T) \equiv {\bf p}_{(o)k} = {\bf p}_{(exp)k},\; k=1,\ldots,N_{(exp)},\\
{\bf p}_{(exp)k} = \frac{N_{(exp)k}}{N_{(exp)}},\; \sum_k N_{(exp)k} = N_{(exp)}, \\
N_{(exp)k} \;{\rm the \; number\; of} \; {\rm events\; having \; index}  \; k \; {\rm in \; the \; ensemble}.
\end{array}
\label{eq:3.1.5.x}
\end{equation}

In eq.~(\ref{eq:3.1.5.x}) \emph{both} the `\emph{identity} ($\equiv$) (referring to quantum \emph{theory})',
as well as the `\emph{equality} ($=$) (referring to the \emph{experimental} measurement process (cf.~(\ref{eq:3.1.3.2.x})))',
should be recognized as two \emph{(different) crucial tools} of `\emph{measurement} within the domain of application of quantum mechanics'.
The present paper is meant to draw the attention to the `\emph{numerical equality} of the probabilities
${\bf p}_{(o)k}$, ${\bf p}_{(a)k}(T)$ and $ {\bf p}_{(exp)k}$ presented in (\ref{eq:3.1.5.x})' because,
by `\emph{accounting for the interaction of the (sub)microscopic object and probe}', this equality is enabling
to reduce to triviality an essential part of the `quantum mechanical measurement problem',
concomitantly opening up new experimental and theoretical possibilities of dealing with `measurement
within the domain of application of quantum mechanics'.\footnote{Although it is an essential part of `measurement
within the quantum mechanical domain', is the (additional) phase-3 \emph{amplification} problem not discussed here because
in general it is dependent on the peculiarities of the measurement procedures (see also \cite{Allahverdyanetal2017},\cite{Nieuwenhuizen2023}).
It is illustrated in \S\ref{subsubsec2.1.1} by the example of the Stern-Gerlach measurement.}

In order to be able to do so it is necessary to change towards a
mathematical representation taking into account that `\emph{what we see' is not the `(sub)microscopic object}' but it rather is a `\emph{measuring instrument's pointer}'. This can\emph{not} be ignored in the way
it has widely been done in the past. On the contrary, in this author's view the \emph{measuring instrument's pointer} has to play a key role
within the domain of application of quantum mechanics. As a matter of fact, whatever we know about the
`(sub)microscopic properties of (sub)microscopic objects' has been learned by observing
`pointers of measuring instruments \emph{having interacted with the objects}' (see \S\ref{subsubsec2.1.1} for an analysis of the Stern-Gerlach
measurement as an example corroborating this observation).

In \S\ref{sec2} some applications of the `quantum mechanical formalism represented by
eqs~(\ref{eq:2.8.10}) through (\ref{eq:3.1.5.x})' are discussed, demonstrating the \emph{necessity} of extending the domain
of application of the quantum mechanical theory so as to encompass POVMs. Thus, in \S\ref{subsubsec2.1.1} the Stern-Gerlach
measurement --during a long time being considered a `paradigm of \emph{standard} quantum mechanics'-- is seen to need
the \emph{generalized} formalism for its description.
In \S\ref{subsec2.2} `joint \emph{non}ideal measurements of \emph{in}compatible observables'
are discussed as `applications of the generalized theory \emph{having experimentally been carried out}' (cf.~\S\ref{subsubsec2.2.1}),
in \S\ref{subsubsec2.2.2} Wigner measures, --although only having a \emph{restricted} applicability--, being referred to as having a possible application in quantum computation.

In \S\ref{sec3} the \emph{physical importance} is discussed of the possibility, offered by the quantum mechanical \emph{theory},
to \emph{equate} the `\emph{experimental} probabilities ${\bf p}_{(exp)k}$ of (\ref{eq:3.1.5.x})' to the \emph{theoretical}
probabilities ${\bf p}_{(a)k}(T)$ and ${\bf p}_{(o)k}$ defined in (\ref{eq:3.1.3.2.x}), the last ones suggesting the
straightforward extension of the set of quantum mechanical observables from PVMs to (P)OVMs\footnote{See
for OVMs \S\ref{subsubsec2.2.2}.} to be found in advanced treatments of
quantum mechanics published during the last 60 years, mainly dealing with probabilities described by POVMs $\{M_{(o)k}(T)\}$
(e.g. \cite{povm}).

The \emph{present} paper is discussing the `\emph{alternative} representation (\ref{eq:3.1.2})'.
Such a discussion is necessary \emph{not only} because of the
`\emph{difference of the physical interpretations} the two expressions are provoking', --(POVMs $\{M_{(o)k}(T)\}$ referring to the \emph{(sub)microscopic object} $(o)$, whereas POVMs $\{M_{(a)k}\}$ are referring to the \emph{measuring instrument/probe} $(a)$)--, but, --perhaps even more importantly--, because the two expressions are
referring to \emph{different times} (to wit, $t=T$ and $t=0$, respectively),
provoking \emph{different interpretations}.
Thus, from the (\emph{ontologically real}) \emph{final} state $\rho_{(a)}(t=T)$ of the \emph{measuring instrument/ancilla}
we are able to obtain `(\emph{epistemologically real}) \emph{information} on the \emph{initial} state $\rho_{(o)}(t=0)$ of the
\emph{(sub)microscopic object}'!

The `\emph{numerical} equality expressed by (\ref{eq:3.1.3.2.x})' is unearthing an `\emph{epistemological} feature'
(already stressed by Bohr (cf.~\S\ref{subsec3.6})) involved in
`quantum mechanics as a theory dealing with \emph{quantitative} properties of \emph{(sub)microscopic} objects',
that feature either \emph{not at all}, or \emph{erroneously}, being discussed in \emph{standard} textbooks as well as in a large part of
the quantum mechanical theoretical literature (among which the author's book \cite{dM2002},
in which --unfortunately-- the difference between an `empiricist \emph{interpretation}' and an `empiricist \emph{approach}'
is not sufficiently stressed (compare \S\ref{subsec3.6} and \S\ref{subsec3.7})).

My conclusion will be that the alleged `\emph{im}possibility of applying the \emph{Schr\"odinger equation} to \emph{quantum measurement}', --the latter equation being replaced by, for instance, von Neumann's well-known \emph{projection postulate} (cf.~\S\ref{subsec3.7})--, is obsolete.
The main purpose of the present paper is to stress the `applicability of the Schr\"odinger equation' to
the `interaction of (sub)microscopic object and probe/ancilla as described by eqs.~(\ref{eq:2.8.10}) and (\ref{eq:3.1.1})',
and to emphasize the \emph{increased usefulness} of the `\emph{generalized} formalism of quantum mechanics'
as compared to the `\emph{standard} one'.

\section{\bf \emph{Generalized} quantum mechanics}
\label{sec2}
`\emph{Generalized} quantum mechanics' is distinguished from `\emph{standard} quantum mechanics' by allowing
\emph{POVMs} as `mathematical entities representing quantum mechanical observables of (sub)microscopic objects'
(rather than restricting attention to PVMs).

In \S\ref{subsubsec2.1.1} the Stern-Gerlach measurement is presented as a
paradigmatic example of the possibility that `\emph{quantum} measurement' need \emph{not} be essentially different from
the well-known examples of `\emph{classical} measurement procedures (like those of temperature and electric current)',
in which `\emph{information} on the object of interest' is obtained by means of `\emph{observation of a change
of a measuring instrument's pointer position}'.
Indeed, far from `just being a paradigm of \emph{standard} quantum measurement' the Stern-Gerlach measurement
turns out to even exhibit the `\emph{experimental feasibility} of \emph{POVMs of generalized quantum mechanics}'
(cf.~\S\ref{subsubsec2.1.2}; see also Martens and de Muynck \cite{MadM92,MadM93,MadM94}, and de Muynck \cite{dM2002}, \S8.3.2).

In \S\ref{subsec2.2} two applications of the `\emph{generalized} quantum formalism' are briefly discussed, pointing to
a considerable enlargement of the latter formalism's `domain of application' as compared to that of the `\emph{standard} one'.
However, it is important to realize that, even though the notion of POVM has its \emph{physical} origin with a realization that
`\emph{within the domain of application of quantum mechanics the probe cannot be left out of the mathematical description
of measurement}', as is expressed by eq.~(\ref{eq:3.1.3.2.x}) it is possible `to cast \emph{experimental} probabilities in terms of the
\emph{(sub)microscopic object $(o)$ only}'.
`\emph{Generalized} quantum mechanics', as defined here, is employing this possibility by allowing POVMs to mathematically represent
`(generalized) quantum mechanical observables of the \emph{(sub)microscopic object $(o)$}',
thus \emph{seemingly} omitting `reference to the \emph{measuring instrument}'.

Unfortunately, `lack of attention to the important role of the \emph{measuring instrument/probe}'
has been continued within the way quantum mechanical probabilities are still dealt with in most textbooks of quantum mechanics,
even if they are allowing POVMs.
It is stressed here, however, that this `\emph{in}dependence from the measuring instrument' is \emph{only an apparent one} since
the `POVM $\{M_{(o)k}(T)\}$ defined in (\ref{eq:3.1.3.2.x})' is `depending on the \emph{interaction of (sub)microscopic object and
probe}'. In order to keep remembering this feature of quantum measurement, `reference to $T$' is being maintained in $M_{(o)k}(T)$.

It should also be stressed that $T$ must be seen as
a symbol referring to the \emph{average} duration of `\emph{individual} measurement processes
represented by the POVM's elements $M_{(o)k}(T), k=1,\ldots,N_{(exp)}$ (cf.~(\ref{eq:3.1.5.x}))
being carried out in a \emph{random} way', the \emph{total} ensemble
having a  duration of the order of magnitude $N_{(exp)}T$ (see also \S\ref{subsec3.3})\footnote{No
`\emph{sub}quantum or hidden-variables theories
being supposed to be able to distinguishing between different members of the ensemble'.}.

\subsection{\bf{The Stern-Gerlach measurement as an application of `generalized quantum measurement'}}
\label{subsec2.1}
Let us, in order to analyze the applicability of the Schr\"odinger equation to quantum measurement,
consider the Stern-Gerlach (SG) measurement \cite{SternGerlach}, which can be looked upon as a paradigmatic example of
`measurement within the domain of application of the quantum mechanical theory' carried out as early as 1922.
In the SG measurement a(n electron) spin of a silver atom, entering an \emph{in}homogeneous static magnetic field
(being produced by a magnet located near the origin $x=y=z=0$) along the $x$-axis, is determined by observation of the point of the atom's impact on a fluorescent screen.

It should be noted that the Stern-Gerlach measurement is a very special example, in the sense that both `object' and `probe' are `aspects of one and the same atom', represented by, respectively, `its spin observable $\sigma_z$' and `its momentum observable $\vec{P}$'.
Hence, in this measurement the `\emph{atom itself}' is playing the role of a `\emph{probe} acting as the pointer of an instrument measuring the \emph{spin} of an electron of the atom', the idea being that there is a unique
relation between the `\emph{initial} direction of the $z$-component $\sigma_z$ of the electron spin' and the
`atom's \emph{final} linear momentum', the latter determining the `position of a visible spot on a fluorescent screen
(situated in the $x$-direction behind the magnet)
caused by the atom's impact' (cf.~\cite{MadM93}).
It is possible to describe this measurement in the quantum mechanical sense discussed in \S\ref{sec1} (as is done in \S\ref{subsubsec2.1.1} (however, see also \S\ref{subsubsec2.1.2}!)).

\subsubsection{\bf{A first approximation}}
\label{subsubsec2.1.1}

In a first approximation of the Stern-Gerlach measurement the `dynamics of the measurement interaction'
can be described by the Schr\"odinger equation (\ref{eq:2.8.10}) in which the Hamiltonian is given according to
(compare \cite{MadM93})
\begin{equation}
 H = \frac{P_x^2+P_y^2+P_z^2}{2m} + (a-bz)\sigma_z,\; a \;{\rm and}\; b \;{\rm constants,}
\label{7a.2.1.0}
\end{equation}
$P_{x_i}=-i\hbar \partial/\partial x_i \; (\{x_i\}=x,y,z)$ being linear momentum observables of the atom,
$\sigma_z$ the `spin observable to be measured', and $a$ and $-bz$, respectively, referring to
position-\emph{in}dependent and -dependent parts of the $z$-component of a time-independent magnetic field the atom
is interacting with.

Since Hamiltonian $H$ (\ref{7a.2.1.0}) is satisfying the equality $[H,\sigma_z]_-=0$,
in this approximation $\sigma_z$ is a `constant of the motion'.
It is seen from Hamiltonian (\ref{7a.2.1.0}) that the atom's \emph{linear momentum} component $P_z$ is
`\emph{not} a constant of the motion'. This latter feature is employed in the Stern-Gerlach measurement as a tool for measuring the
`pre-measurement spin component $\sigma_z$ of (an electron of) the atom'.
As observed by Martens and de Muynck \cite{MadM93} (also \cite{dM2002}, \S8.3.2),
in the Heisenberg picture of the present approximation we find (since in the experiment $P_z(0)=0$)
\begin{equation}
P_z(t) = \frac{1}{2}b\sigma_z t,\; t\leq \tau
\label{7a.2.1.3}
\end{equation}
(cf.~\cite{MadM93},~eq.~(3))\footnote{It should be reminded that eq.~(\ref{7a.2.1.3}) is the `\emph{Heisenberg} part' of the
quantity $<\psi|P_z(t) - \frac{1}{2}b\sigma_z t|\psi>,\; |\psi>$ being the `state of the atom (encompassing \emph{both} the atom's momentum as well as the electron spin)'.}, $\tau$ being the time at which the atom is leaving the `region of space in which the magnetic-field is non-zero',
and, hence, $P_z(t)=P_z(\tau)$ for $\tau\leq t\leq T$, $T$ the time at which the atom is reaching the fluorescent screen.
Hence, according to this account there is a \emph{strict} correlation between the
`\emph{initial} values $\sigma_z = \pm$ of the (electron) spin of the atom' and the `\emph{final} values ($\pm |p_z|$) of $P_z$',
thus allowing `\emph{post}-measurement spots in the \emph{upper} ($z>0$) part of the screen' to be interpreted as
`\emph{pre}-measurement values $\sigma_z = +$ ',
and in the \emph{lower} ($z<0$) part as $\sigma_z = -$ (or vice versa (depending on the sign of parameter $b$)).
The measurement results of this approach to the Stern-Gerlach measurement can finally be recorded as (compare
(\ref{eq:3.1.5.x}))
\begin{equation}
\begin{array}{l}
 {\bf p}_{(o)}(\sigma_z=+, t=0) \equiv {\bf p}_{(a)(P_z(\tau) >0)}:= {\bf p}_{(exp)+},\\
 {\bf p}_{(o)}(\sigma_z=-, t=0) \equiv {\bf p}_{(a)(P_z(\tau) <0)}:= {\bf p}_{(exp)-}, \\
 {\rm or} \;  \pm \rightarrow \mp \; {(\rm {depending \;on\; the\; sign\; of\; parameter}}\; b),\\
  {\bf p}_{(exp)+} + {\bf p}_{(exp)-} = 1,
\end{array}
\label{7a.2.1.4}
\end{equation}
$N_{(a)(P_z(\tau)>0)}$ resp. $N_{(a)(P_z(\tau)<0)}$ being the \emph{experimentally obtained}
`numbers of probes $(a)$ during the \emph{total} measurement time $N_{(a)}T$
($N_{(a)}$  being the \emph{total} number of impacts on the \emph{whole} screen)'.

For sufficiently large values of $N_{(a)}$ `solutions of the Schr\"odinger equation (\ref{eq:2.8.10}) with Hamiltonian (\ref{7a.2.1.0})'
are in reasonable (but not perfect) agreement with these results.
Indeed, as will be seen in \S\ref{subsubsec2.1.2}, applicability of equation (\ref{eq:2.8.10})
is thwarted because Hamiltonian (\ref{7a.2.1.0}) is only yielding an \emph{approximate} representation of the `interaction of the atom
with a magnetic field', a (more) correct one
\emph{not} yielding the `\emph{strict} correlation of observables $\sigma_z$ and $P_z(t)$ being assumed in (\ref{7a.2.1.3})'.

Postponing till \S\ref{subsec3.4} a discussion of this latter feature of the Stern-Gerlach measurement,
in a \emph{single} execution of the measurement it is reasonable to interpret the `appearance of a (single) spot on the screen'
as an \emph{event} in which `at $t=T$ the ancilla ($a$) is at its \emph{post-measurement} pointer position'.
The probabilistic aspect of this measurement is described by eq.~(\ref{eq:3.1.2}), the quantities ${\bf p}_{(a)k}(T),\; k=\pm$ yielding
`\emph{statistical} information on the relative frequencies of the latter events'.
However, our interest is \emph{not} directed toward the `\emph{final} state of the \emph{ancilla}', but instead toward the
`\emph{initial} state of the (sub)microscopic object', probabilities ${\bf p}_{(a)k}(T)$ being equal to probabilities ${\bf p}_{(o)k}$
(cf.~(\ref{eq:3.1.3.2.x}) and (\ref{eq:3.1.5.x})).\\

\subsubsection{\bf{Correcting the quantum mechanical description of the Stern-Gerlach measurement}}
\label{subsubsec2.1.2}
It was demonstrated by Scully, Lamb Jr. and Barut \cite{ScLaBa87} that Hamiltonian
(\ref{7a.2.1.0}) is wanting, and does \emph{not exactly} describe the Stern-Gerlach measurement.
As a matter of fact, Hamiltonian (\ref{7a.2.1.0}) is \emph{un}physical because
the interaction term $a-bZ$ is \emph{in}consistent with the requirement that the magnetic field
satisfy the condition ${\bf \nabla}\cdot {\bf B}=0$: that condition can\emph{not} be satisfied
if the (\emph{in}homogeneous) magnetic field has \emph{only one} component (as is assumed in eq.~(\ref{7a.2.1.0})).

It is possible (cf. Martens and de Muynck \cite{MadM93}, also \cite{dM2002}, \S8.3.2)
to deal with this objection by correcting the magnetic field so as to satisfy the above-mentioned condition.
This can be done by changing the Hamiltonian of the `Schr\"odinger equation describing the Stern-Gerlach measurement'
according to
\begin{equation}
H=\frac{P_x^2+P_y^2+P_z^2}{2m} +\frac{\mu}{2} \{bY\sigma_y + (a-bZ)\sigma_z\}. \label{7a.2.1}
\end{equation}
It should be noted that, since $[\sigma_z,\sigma_y]_- \neq 0$, $\sigma_z$ is \emph{no longer} a constant of the motion.
As a matter of fact,
the `\emph{interaction} of (sub)microscopic object and probe' (as described by eq.~(\ref{eq:3.1.1}))
is influencing the measurement result of $\sigma_z$, thus
\emph{preventing} the Stern-Gerlach measurement from being a(n \emph{ideal}) measurement of the latter
observable.\footnote{See also \S\ref{sec3} for a discussion of this issue.}

However, with Hamiltonian (\ref{7a.2.1}) observable $L_x-\sigma_x/2$ \emph{is} a constant of the motion.\footnote{I here ignore
the possibility of redefining $L_x-\sigma_x/2$ as
$L_x + \sigma_x$, thus enabling an interpretation of this conservation law as `conservation of \emph{total} angular momentum $J_x$ '.}
Moreover, if a corresponding symmetry is satisfied by the `spatial part of the initial state of the
atom/probe',\footnote{Accepted here without discussion (cf.~Martens and de Muynck \cite{MadM93}, also \cite{dM2002}, \S8.3).}
then the Stern-Gerlach measurement is still a measurement of $\sigma_z$, be it a \emph{non}-ideal one (cf.~\cite{MadM93},~\S3),
$ M_{(o)k}(T) $ (as defined in (\ref{eq:3.1.3.2.x})) satisfying
\begin{equation}
 M_{(o)k}(T) = \sum_{k'=\pm} \lambda_{kk'}E_{(o)k'}, \; k=1,\ldots,N,\; E_{(o)k'} = |(o)k'\rangle\langle (o)k'|,\; k'=\pm,
\label{7a.2.1.2}
\end{equation}
$N$ the `number of \emph{different} pointer positions' (which may be \emph{different}
from the `number of eigenvalues of the atom's spin observable').
The quantities $\lambda_{kk'}$ are elements of a so-called `\emph{stochastic} matrix' $\{\lambda_{kk'}\}$.
Note that, contrary to a \emph{mathematical} custom, \emph{within physics} usually the `\emph{left} stochastic matrix' is used.
Note also that, in order to be within the domain of application of quantum mechanics,
`relative frequencies $N_{(exp)k}/N_{(exp)}$ of individual measurement results $k$' must have `large-$N$ limits',
thus requiring quantum mechanical stochasticity to satisfy the `law of large numbers'.
Physicists usually are referring to this matrix as a `\emph{non}ideality matrix'
(\emph{not} necessarily square\footnote{Restricting ourselves here to $N=2$, the (non)ideality matrices are square.}),
to be expressed in quantum mechanical terms according to
\begin{equation}
\begin{array}{l}
\lambda_{kk'}=Tr_{(o)} M_{(o)k}(T) E_{(o)k'}, \lambda_{kk'} \;\mbox{\rm real and} \geq 0, k=1,\ldots,N,\;
\sum_{k=1}^{N} \lambda_{kk'}=1,\; k'=\pm.
\end{array}
\label{7a.2.1.1xx}
\end{equation}
A measurement described by a POVM like (\ref{7a.2.1.2}) is referred to as a `\emph{non}-ideal measurement of PVM $\{E_{(o)k'}\}$',
its measurement probabilities being given, in agreement with (\ref{eq:3.1.3.2.x}) and (\ref{eq:3.1.5.x}), by
\begin{equation}
\begin{array}{l}
{\bf p}_{(o)k} = \sum_{k'=\pm} \lambda_{kk'} Tr_{(o)}(\rho_{(o)}(t=0) E_{(o)k'}),\; k= 1,\ldots,N. \\
\end{array}
\label{7a.2.1.1}
\end{equation}
\noindent Restricting ourselves to $N({\cal H}_{(o)})$ $ = 2$, the `\emph{measured} relative frequencies ${\bf p}_{(exp)k},\; k= \pm$
corresponding to POVM (\ref{7a.2.1.2}) (obtained in an `\emph{ensemble} of measurements carried out
on identically prepared \emph{individual} atoms' by observing the positions of the points of impact of the
atoms on the screen) are \emph{not} equal to the \emph{standard} probabilities (\ref{eq:3.1.0}).

However, if the inverse $\{\lambda^{-1}_{k'k}\}$ of the matrix $\{\lambda_{kk'}\}$ exists, then the `\emph{standard} probabilities can be obtained by means of the equality
$Tr_{(o)}(\rho_{(o)}(t=0) E_{(o)k'})= \sum_{k=1}^N \lambda^{-1}_{k'k} {\bf p}_{(o)k} \; ,k'=\pm$
 (see also \S\ref{subsubsec2.2.2})\footnote{An example
can be found in \cite{dM2002}, \S7.2.1.}.
Contrary to the result found in \S\ref{subsubsec2.1.1}, the \emph{strict} correlation between the pre-measurement value
of $\sigma_z$ and the post-measurement value of $P_z$ expressed by (\ref{7a.2.1.4}) is \emph{not} found.
Evidently, in agreement with \cite{ScLaBa87} it is seen that the Stern-Gerlach measurement is `\emph{not} the \emph{ideal} measurement of
the spin observable which during a long time it was thought to be'.
The \emph{nonideality} matrix $\{\lambda_{kk'}\}$ defined in (\ref{7a.2.1.1xx})
(its elements $\lambda_{kk'}$ being explicitly displayed in~\cite{MadM93})
is reflecting the `\emph{deviation} from ideality' realized
by changing the Hamiltonian from (\ref{7a.2.1.0}) into (\ref{7a.2.1}), \emph{ideality} to be realized if $\lambda_{kk'}=\delta_{kk'}$.
Note that in case of \emph{non}ideality the stochastic matrix (\ref{7a.2.1.1xx}) needs \emph{not} be a \emph{square}
one, the number of rows possibly differing from the number of columns (hence, a \emph{non}ideal spin-$1/2$
measurement is possible, e.g. \cite{dM2002}, eq.~(7.36)).

It should be stressed that the POVM $\{M_{(o)k}(T)\}$ defined by (\ref{7a.2.1.2}), and, hence,
the \emph{stochastic} matrix $\{\lambda_{kk'}\}$ as well as the PVM $\{E_{(o)k'}\}$, are \emph{theoretically derived}\footnote{Due to a necessity of applying perturbation theory often in an approximate way.} from the `Schr\"odinger equation describing the \emph{measurement process}',
thus expressing a close connection of quantum mechanics to stochasticity (e.g. Beyer and Paul \cite{stoch}).

It should also be stressed that in the `derivation of eqs~(\ref{7a.2.1.1xx}) and (\ref{7a.2.1.1}) thus obtained' use is made
of the \emph{numerical} equality of probabilities (\ref{eq:3.1.3.2.x}) and (\ref{eq:3.1.5.x},\ref{7a.2.1.2}),
thus enabling to \emph{equate} a(n `\emph{observed}) \emph{post}-measurement probability of the \emph{probe~(a)}'
(i.e.~(\ref{eq:3.1.5.x},\ref{7a.2.1.2})) to an `(\emph{un}observed) \emph{pre}-measurement probability of the
\emph{(sub)microscopic object $(o)$} (i.e. (\ref{eq:3.1.3.2.x}))'.

Relegating to \S\ref{sec3} a discussion of the above-mentioned equality in terms of the `\emph{difference} between \emph{ontological} and
\emph{epistemological} meanings of the quantum mechanical formalism', it is stressed already here that --often \emph{erroneously}--
the relation between $Tr_{(o)} \rho_{(o)}(t=0) E_{(o)k'}$ and
${\bf p}_k= Tr_{(o)} \rho_{(o)}(t=0) M_{(o)k}(T)$ is interpreted as a relation between
`\emph{pre- and} \emph{post-}measurement probabilities of the \emph{(sub)microscopic object $(o)$}', probabilities ${\bf p}_k$
being \emph{supposed} to be `\emph{post-measurement properties of the (sub)microscopic object $(o)$ being disturbed by the measurement}'
rather than `\emph{post-measurement properties of the probe $(a)$}',
the \emph{latter difference} being the subject of the present paper.
However, the `\emph{disturbance} interpretation referred to here'
is neglecting that the `interaction of (sub)microscopic object and probe' has \emph{two different} consequences, viz.
a \emph{determinative} and a \emph{disturbance} one, the former referring to a `measurement's influence on the \emph{probe}',
the second one to its `influence on the \emph{(sub)microscopic object}'.
Unfortunately, due to within the quantum mechanical formalism `\emph{avoiding reference
to the probe}',\footnote{In \S\ref{subsec3.4} this custom is referred to as `sticking to the \emph{classical} paradigm'.}
this difference has during a long time remained largely \emph{un}noticed, thus generating lots of
confusion (cf.~\S\S\ref{subsec3.6} and \ref{subsec3.7}).

Perhaps this is a good place to bestow some attention on the issue of `\emph{calibration} of the measuring instrument'.
Let us take the Stern-Gerlach measurement as a simple example.
Probabilities ${\bf p}_{(exp)k}, k= \pm$ (as defined in (\ref{7a.2.1.4}))\footnote{Higher spin values are not discussed here.}
being experimentally found to be different from $1$ c.q. $0$,
in this measurement an `\emph{individual} screen spot caused by the impact of a probe'
does not tell us whether it should be counted as eigenvalue $\sigma_z = +$ or as $\sigma_z = -$.
As was seen above, by the interaction with the magnetic field the probe could have been sent to the ``wrong''
part of the screen, thus representing a certain \emph{non}-ideality of the Stern-Gerlach measurement
(described by a deviation of the non-ideality matrix $\{\lambda_{kk'}\}$ from $\{\delta_{kk'}\}$).

However, on the basis of the \emph{statistical} theory of the `interaction of (sub)microscopic object ($o$) and probe ($a$) as represented by (\ref{eq:3.1.1})' quantum mechanics is able to deal with this problem,
be it in a \emph{statistical} sense only. Thus, by for the Stern-Gerlach problem
carrying out `\emph{calibration} measurements' by starting from the eigenstates $|\sigma_z = +\rangle$ c.q. $|\sigma_z = -\rangle$
as `\emph{initial} states of the (sub)microscopic object ($o$)',
it can be seen to what extent probabilities ${\bf p}_{(exp)k}, k= \pm$ (as defined in (\ref{7a.2.1.4})) are \emph{different
from} $1$ c.q. $0$, the differences being represented by the nonideality matrix $\{\lambda_{kk'}\}$
(presented in (\ref{7a.2.1.1xx})) according to $\lambda_{++} ={\bf p}_{(exp)+},\; \lambda_{--} = {\bf p}_{(exp)-}$,
$\lambda_{-+} =1 - {\bf p}_{(exp)+},\; \lambda_{+-} =1 - {\bf p}_{(exp)-}$.

It is important to realize that, although the nonideality matrix $\{\lambda_{kk'}\}$ is \emph{experimentally} found
from the \emph{two} calibration measurements referred to above, it is seen from (\ref{7a.2.1.1xx}) that its elements are
`\emph{in}dependent from the \emph{initial} state $\rho_{(o)}(t=0)$'. Hence, when employing the \emph{same} measurement arrangement,
the \emph{same} nonideality matrix can be used \emph{independently} of which `\emph{initial state} of object $(o)$' has been prepared.\\

\subsubsection{\bf{The Stern-Gerlach measurement as a `\emph{joint} nonideal measurement of two \emph{in}compatible standard observables'}}
\label{subsubsec2.1.3}
A third version of the Stern-Gerlach measurement is considered here, in which the magnetic field has
a \emph{quadrupolar} configuration, to be obtained by taking in (\ref{7a.2.1}) $a=0$, thus yielding
\begin{equation}
H_{(a=0)}=\frac{P_x^2+P_y^2+P_z^2}{2m} +\frac{\mu}{2}\{bY\sigma_y -bZ\sigma_z\}.
\label{7a.2.3.2}
\end{equation}
By Martens and de Muynck \cite{MadM93} (also \cite{dM2002}, \S8.3.3) this measurement, too, was
analyzed in terms of `\emph{interaction} of the (sub)microscopic object (i.e. the atom's electron spin) and the probe
(the latter being represented by the atom's \emph{linear momentum} observable)'.\footnote{See \S\ref{subsec3.4} for a discussion
of Hamiltonian (\ref{7a.2.3.2}) as describing a `measurement of the atom's spin'.}
By requiring `the spatial part of the initial state of the atom to satisfy $L=0$'
a generalization of (\ref{7a.2.1.2},\ref{7a.2.1.1xx}) was obtained, to the effect that
this measurement is able to yield `\emph{joint},\footnote{Usually, in a less accurate way being referred to as `simultaneous'.}
though \emph{non}ideal, information on the \emph{two} \emph{in}compatible spin observables $\sigma_y$ and $\sigma_z$ '.
Thus, analogously to \S\ref{subsubsec2.1.2} applying the \emph{numerical} equality expressed by (\ref{eq:3.1.3.2.x}), a
\emph{bi}variate POVM $\{M_{(o)k\ell}(T)\}$ was found ((cf.~\cite{MadM93}, \cite{dM2002}, \S8.3.3), \cite{Buschetal2013})
generalizing (\ref{7a.2.1.2}) so as to satisfy the following relations
 \begin{equation}
\begin{array}{l}
\sum_{\ell=1}^{\tilde{N}} M_{(o)k\ell}(T) = \sum_{k'} \lambda_{kk'} E_{(o)k'}, \; k= 1,\ldots, N,\; E_{(o)k'} =
|(o)k'\rangle\langle (o)k'|,\; k'= \pm,\\
\sum_{k=1}^{N} M_{(o)k\ell}(T) = \sum_{\ell'} \mu_{\ell\ell'} F_{(o)\ell'}, \; \ell= 1,\ldots, \tilde{N},\;
F_{(o)\ell'}= |(o)\ell'\rangle\langle (o)\ell'|,\; \ell'= \pm,\\
|(o)k'\rangle \; {\rm and} \; |(o)\ell'\rangle \;{\rm  eigenvectors\; of}\; \sigma_{y} \;{\rm and} \; \sigma_{z},\;{\rm respectively},
\end{array}
\label{7.9.1}
\end{equation}
POVMs $\{\sum_{\ell=1}^{\tilde{N}} M_{(o)k\ell}(T)\}$ and $\{\sum_{k=1}^{N} M_{(o)k\ell}(T)\}$
corresponding to \emph{non}ideal measurements of PVMs $\{E_{(o)k'}\}$ and $\{F_{(o)\ell'}\}$, respectively,
$N$ and $\tilde{N}$ not necessarily being equal.
Both $\{\lambda_{kk'}\}$ and $\{\mu_{\ell\ell'}\}$ are \emph{stochastic} matrices, both satisfying (\ref{7a.2.1.1xx}), in (\ref{7.9.1})
representing, respectively, the `\emph{non}idealities
of measurement results of \emph{in}compatible observables $\sigma_y$ and $\sigma_z$ \emph{if measured jointly}', \emph{bi}variate
experimental probabilities being given by ${\bf p}_{k\ell} = Tr_{(o)} \rho_{(o)}(t=0) M_{(o)k\ell}(T)$.\footnote{Once again
$\rho_{(o)}(t=0)$ is referring to the `\emph{initial} state of the \emph{(sub)microscopic object}
(in the present example this object being represented by the atom's `\emph{spin} angular momentum components')'.}

In de Muynck \cite{dM2002}, Chapt.~8 (see also de Muynck \cite{dM98,dM2000})
a number of \emph{experimentally realized} measurements satisfying (\ref{7.9.1}) has been reviewed,\footnote{Amongst which those gaining
Haroche \cite{Harochecomplementarity2001,Harochecompl2001} the (shared) 2012 Nobel Prize in Physics
``for ground-breaking experimental methods that enable measuring and manipulations of individual quantum systems''.}
demonstrating the physical applicability of the \emph{generalized} quantum mechanical formalism, as well as
the \emph{experimental} feasibility of `\emph{joint nonideal} measurements of \emph{in}compatible standard observables'.

It, perhaps, is important to warn already here \emph{against} an interpretation of eqs~(\ref{7.9.1}) as `\emph{in a realist sense}
referring to the (sub)microscopic object ($o$) as --be it in a probabilistic sense-- in the \emph{initial}
state $\rho_{(o)}(t=0)$ \emph{simultaneously} ``\emph{having}'' values of \emph{both} standard observables
$\{E_{(o)k'}\}$ and $\{F_{(o)\ell'}\}$'.
As stressed above, even though being \emph{mathematically} expressed as `\emph{properties of the (sub)microscopic object}', the two POVMs
$\{\sum_{\ell=1}^{\tilde{N}} M_{(o)k\ell}(T)\}$ and $\{\sum_{k=1}^{N} M_{(o)k\ell}(T)\}$ are \emph{physically} referring
to `\emph{properties of measuring instruments}' (see \S\ref{subsec3.4} for a discussion of this issue).

\subsection{\bf{Applications of \emph{generalized} quantum mechanics}}
\label{subsec2.2}

In the present subsection I restrict myself to the possibility --yielded by the \emph{numerical} equality of the properties defined in
eq.~(\ref{eq:3.1.3.2.x})-- of interpreting a `\emph{post}-measurement probability of the \emph{probe}'
as a `\emph{pre}-measurement probability of the \emph{(sub)microscopic object}'.
Nowadays the idea of a POVM (rather than a PVM) as a `mathematical representation of a quantum mechanical observable'
seems to have been generally accepted, be it often as `just a straightforward \emph{mathematical} generalization
of the notion of a quantum mechanical observable (viewed upon as a property of the \emph{(sub)microscopic object} ($o$)',
\emph{without} recognizing its \emph{physical} meaning as as a `property of the ancilla $(a)$').
Unfortunately, due to restricting POVMs
--analogously to PVMs-- to
`properties of the \emph{(sub)microscopic object} ($o$)', the possibility of an `interpretation of probabilities
${\bf p}_{(o)k}$ as probabilities ${\bf p}_{(a)k}(T)$ of the \emph{measuring instrument/probe} $(a)$)',
offered by (\ref{eq:3.1.3.2.x}), has been largely neglected (see \S\ref{sec3} for a discussion of
the disturbing consequences of neglecting this issue).\footnote{In the following I shall maintain in POVMs
`\emph{explicit} reference to \emph{measurement} time $T$ (e.g. (\ref{7a.2.1.2}))'
whenever their `\emph{physical} meaning as a property of the \emph{object}' is intended.}

It is the main purpose of the present paper to stress the `\emph{ontological} role of the \emph{measuring instrument}'
which --although present in the `\emph{empiricist} interpretation of the quantum mechanical formalism
entertained in \cite{dM2002}'-- was too much concealed there.
Relegating to \S\ref{sec3} a discussion of different views on this subject, it is necessary to realize that
the `human observer is \emph{un}able to see the (sub)microscopic object'.
Measuring instruments are necessary to bridge the gap between that object and the observer.
At best, the human observer is able to see the `\emph{pointer positions of his measuring instrument}'
\emph{after a certain amplification has been realized}.\footnote{Compare the Stern-Gerlach measurement discussed in \S\ref{subsec2.1}
for an example how such amplification can be physically realized.}

Note that this issue is not evident from the extension of quantum probabilities
to `expectation values of POVMs' ${\bf p}_{(o)k} = Tr_{(o)} \rho_{(o)}(t=0)M_{(o)k}(T)$
(rather than the PVMs of the first half of the $20^{th}$ century).
It, however, is expressed by the `\emph{numerical} equality
of ${\bf p}_{(a)k}(T)$ and ${\bf p}_{(o)k}$ stressed in (\ref{eq:3.1.3.2.x})',
explicitly referring to the `\emph{final pointer positions of the measuring instrument/probe}', and
having been overlooked in most of the quantum physical literature of the $20^{th}$ century (see also \S\ref{subsec3.4}).\\

\subsubsection{\bf{Ideal and \emph{non}ideal measurements of \emph{standard} observables}}
\label{subsubsec2.2.1}

As observed in \S\ref{subsec2.1}, the `\emph{(sub)microscopic} part of the Stern-Gerlach measurement'
can be dealt with essentially in \emph{quantum mechanical} terms.
The importance of referring here to this example is that it is illustrating a feature of `measurement within the \emph{(sub)microscopic}
domain' allowing measurement probabilities
to \emph{deviate} from the ones taken into account by `\emph{standard} quantum mechanics'.
The present paper is dealing with a \emph{generalization} of eq.~(\ref{eq:3.1.0}),
to the effect that the `(standard) \emph{projection}-valued measure (PVM) $\{E_{(o)j}\}$' considered in that equation is
replaced by a POVM
\begin{equation}
M_{(o)k}(T)= \sum_{k'=1}^{N({\cal H}_{(o)})} \lambda_{kk'} E_{(o)k'},\;k=1,\ldots, N, \;
\forall_{k'} \sum_{k=1}^N \lambda_{kk'}= 1,
\label{7a.2.1.1x}
\end{equation}
representing a `\emph{non-ideal} version of a \emph{standard} measurement'
(note that the `number $N$ of available pointer positions' may be different from the dimension $N({\cal H}_{(o)})$
of the Hilbert space of possible eigenvectors of object $(o)$,
thus allowing nonideality matrices $\{\lambda_{kk'}\}$ to be \emph{non}-square).

In the literature of mathematical physics many so-called `nonideality measures' can be found (e.g.
McEliece, \cite{McEl2004}, Chapt.~1 and App.~B, also \cite{dM2002}, \S7.8),
the `average row entropy of the nonideality matrix $\{\lambda_{kk'}\}$',
\begin{equation}
J(\{\lambda_{kk'}\}) := - \frac{1}{N} \sum_{k=1}^{N} \sum_{k'=1}^{N({\cal H}_{(o)})}
\lambda_{kk'} \ln (\frac{\lambda_{kk'}}{\sum_{k''=1}^{N({\cal H}_{(o)})} \lambda_{kk''}})
\label{7.8.6}
\end{equation}
(cf.~\cite{MadM92}, eq.~(10); also  \cite{dM2002}, p.~374) being a particularly useful measure of the `\emph{non}ideality of a
measurement represented by a POVM $\{M_{(o)k}(T)\}$ as defined by (\ref{7a.2.1.1x})'.
It is important to note that $J(\{\lambda_{kk'}\}) \geq 0$, and is vanishing if and only if $\lambda_{kk'} =\delta_{kk'}$
(i.e. in case of an `\emph{ideal} measurement of a \emph{PVM} $\{E_{(o)k'}\}$').
It should also be noted that, although
--generalizing (\ref{7a.2.1.1xx})--
$\sum_{k=1}^{N}\lambda_{kk'} =1$, in general $\sum_{k'=1}^{N({\cal H}_{(o)})}\lambda_{kk'} \neq 1$
(\emph{no double} stochasticity!).\footnote{Note that this is  \emph{generalizing}
the way this theory is used, for instance, by Nielsen and Chuang \cite{NielsenChuang}, p.~511/513
(the latter \emph{requiring double} stochasticity).}

The measure $J(\{\lambda_{kk'}\})$ (as given by (\ref{7.8.6})) has been obtained from the `mutual information'
(cf.~\cite{MadM90a}, p.~279)
by \emph{assuming} $\forall_k \;{\bf p}_k= 1/N({\cal H}_{(o)})$, thus \emph{confirming} the `\emph{in}dependence of $J(\{\lambda_{kk'}\})$ from
the \emph{initial} density operator $\rho_{(o)}(t=0)$ of the (sub)microscopic object'. This measure evidently is taking into account
{\em only} the `uncertainty generated by \emph{the measurement process}'.
It should be realized, however, that `\emph{preparation of the initial state of the (sub)}-\emph{microscopic object}'
is an `\emph{additional} source of \emph{un}certainty', this latter source being symbolized by the density operator $\rho_{(o)}(t=0)
= \sum_{k=1}^{N({\cal H}_{(o)})} r_k |r_k\rangle\langle r_k|,\; r_k $ the eigenvalues of $\rho_{(o)}(t=0)$.
Representing the `uncertainty of \emph{preparation} of the initial state $\rho_{(o)}(t=0)$ of the (sub)microscopic object'
by the `von Neumann entropy' (e.g. \cite{dM2002}, \S1.4.1)
\begin{equation}
H_{(vN)}(\rho_{(o)}(t=0)) = -\sum_{k=1}^{N({\cal H}_{(o)})} r_k\log r_k,
\label{2.2.1.xx}
\end{equation}
($r_k$ satisfying $r_k \geq 0$, $\sum_{k=1}^{N({\cal H}_{(o)})}r_k =1$),
the `\emph{total} uncertainty $\Delta(\rho_{(o)}(t=0), \{M_{(o)k}(T)\})$ of the measurement process' --dealing with
`\emph{both preparation and measurement}'-- can be defined according to
\begin{equation}
\Delta(\rho_{(o)}(t=0), \{M_{(o)k}(T)\}):= H_{(vN)}(\rho_{(o)}(t=0)) + J(\{\lambda_{kk'}\}).
\label{2.2.1.1}
\end{equation}
Since $H_{(vN)}(\rho_{(o)})\geq 0$ for \emph{any} density operator $\rho_{(o)}$, we obtain the inequality
\begin{equation}
\Delta(\rho_{(o)}(t=0), \{M_{(o)k}(T)\})\geq J(\{\lambda_{kk'}\}).
\label{2.2.1.2}
\end{equation}

Once again generalizing the results obtained for the Stern-Gerlach measurement (cf.~\S\ref{subsubsec2.1.3}),
and taking into account that two \emph{different} probes may be present, then
--generalizing (\ref{7.9.1}) (but, as above, restricting ourselves to \emph{non}-degenerate standard observables)--
a mathematical definition of an `\emph{observable} symbolizing a
\emph{joint nonideal} measurement of \emph{two} \emph{in}compatible standard observables $\{E_{(o)k}\}$ and $\{F_{(o)\ell}\}$'
may be represented by a \emph{bi}variate POVM $\{M_{(o)k\ell}(T)\}$ satisfying
\begin{equation}
\begin{array}{l}
\sum_{\ell=1}^{\tilde{N}} M_{(o)k\ell}(T) = \sum_{k'= 1}^{N({\cal H}_{(o)})} \lambda_{kk'} E_{(o)k'}, \; k= 1,\ldots,N,\;
E_{(o)k} = |(o)k\rangle\langle (o)k|,\; k= 1,\ldots,N({\cal H}_{(o)}),\\
\sum_{k=1}^{N} M_{(o)k\ell}(T) = \sum_{\ell'= 1}^{{\tilde{N}}{(\cal H}_{(o)})} \mu_{\ell\ell'} F_{(o)\ell'}, \;
\ell=1,\ldots,\tilde{N},\;
F_{(o)\ell}= |(o)\ell\rangle\langle (o)\ell|,\;\ell=1,\ldots,\tilde{N}({\cal H}_{(o)})),
\end{array}
\label{7.9.1xxx}
\end{equation}
stochastic matrices $\{\lambda_{kk'}\}$ and $\{\mu_{\ell\ell'}\}$
generalizing the ones defined in (\ref{7.9.1}), $N$ and $\tilde{N}$ \emph{not} necessarily being equal, hence
allowing in (\ref{7.9.1xxx}) $(N, N({\cal H}_{(o)}))$ to be different from $(\tilde{N}, \tilde{N}({\cal H}_{(o)})$.

Assuming `observations of the two probes' to be \emph{distinguishable} processes, we may define for
each of the `two POVMs defined in (\ref{7.9.1xxx})' nonideality measures analogous to (\ref{7.8.6}),
and accept their sum $J(\{\lambda_{kk'}\}) + J(\{\mu_{\ell\ell'}\})$ as a `measure of the \emph{non}-ideality of the \emph{joint} nonideal
measurement processes represented by POVM $\{M_{(o)k\ell}(T)\}$'.
Generalizing (\ref{2.2.1.2}) to the `\emph{joint nonideal} measurement corresponding to POVM $\{M_{(o)k\ell}(T)\}$',
we then finally obtain the inequality
\begin{equation}
\Delta(\rho_{(o)}(t=0), \{M_{(o)k\ell}(T)\})\geq J(\{\lambda_{kk'}\}) + J(\{\mu_{\ell\ell'}\}),
\label{2.2.1.3}
\end{equation}
to be referred to as the `\emph{generalized} Martens inequality',
which can be seen as the `uncertainty inequality' satisfied by a `\emph{joint non}ideal measurement of standard observables
$\{E_{(o)k'}\}$ and $\{F_{(o)\ell'}\}$', to be contrasted with the `Heisenberg inequality' (see \S\ref{subsec3.5} for a discussion of this issue).
The `generalized Martens inequality' (\ref{2.2.1.3}) should be contrasted with the `Martens inequality'
\begin{equation}
J(\{\lambda_{ii'}\}) + J(\{\mu_{jj'}\}) \geq -\ln(\max_{ij}(Tr_{(o)} E_{(o)i} F_{(o)j})
\label{eq:3.1.9}
\end{equation}
(cf.~Martens and de Muynck~\cite{MadM90b}, p.~365, def.~7 (also \cite{dM2002}, \S7.10.2), (7.106),
which was intended to describe `\emph{mutual} disturbance of measurement results
in a joint \emph{non}ideal measurement of \emph{in}compatible standard observables' as defined by (\ref{7.9.1})),
this inequality being thought to be comparable to `Heisenberg's uncertainty relation' (cf.~(\ref{eq:3.1.10})).

However, this latter idea is \emph{only partly} correct.
The `Martens inequality (\ref{eq:3.1.9})', although mathematically correct, does \emph{not} have the \emph{physical} relevance
which it should have in order to be comparable with the `Heisenberg inequality'.
As a matter of fact, contrary to inequality  (\ref{2.2.1.3})
the `Martens inequality' is \emph{only} referring to the `\emph{measuring instrument(s)}', whereas the `Heisenberg inequality'
is \emph{also} referring to the `(preparation of the) \emph{initial state} of the (sub)microscopic object'!
Inequality (\ref{2.2.1.3}) is improving on this predicament.\\

\subsubsection{\bf{`Wigner measure' versus `polar/Schmidt decomposition'; why \emph{only the first one} is physically applicable}}
\label{subsubsec2.2.2}

In this subsection two applications of the mathematical formalism of quantum mechanics are contrasted with respect to
their physical applicability. It will be seen that --contrary to appearances-- `Wigner measures' have a certain physical
applicability, whereas the `polar/Schmidt decomposition' does \emph{not} have one.\\

{\bf Wigner measures, and their (restricted) \emph{physical} applicability.}

Wigner measures may be obtained from POVMs $\{M_{(o)k}(T)=\sum_{k'=1}^{N(H_{(o)})} \lambda_{kk'}E_{(o)k'}\}$ (\ref{7a.2.1.2})
of `\emph{non}ideal measurements of a PVM $\{E_{(o)k}\}$, ${\bf p}_{(o)k} = Tr_{(o)}\rho_{(o)}(t=0) M_{(o)k}(T),\; k= 1,\ldots, N$
(representing the probabilities (\ref{7a.2.1.1})
of a measuring instrument's final \emph{pointer} positions)'. A \emph{Wigner measure} may be defined as an operator-valued measure
(OVM)\footnote{cf.~\cite{dM2002}, \S7.9.3, also l.c. Appendices A.12.2 and A.12.3.} $\{W_{(o)k'}\},\; W_{(o)k'} =$
$\sum_{k} \lambda^{-1}_{k'k}E_{(o)k}$ (the matrix $\{\lambda^{-1}_{k'k}\}$ being the \emph{inverse} of
$\{\lambda_{kk'}\}$)\footnote{See \cite{dM2002}, \S7.6.3 for the question of existence of these inverses.}
corresponding to ``probabilities'' $Tr_{(o)}\rho_{(o)}(t=0) W_{(o)k'}$ that may have \emph{negative} values
(cf.~Wigner \cite{Wig32}) since $\lambda^{-1}_{k'k}$ may be negative.

Let $\{M_{(o)k\ell}(T)\}$ be the (bi-variate) POVM of a `\emph{joint} \emph{non}ideal measurement of two PVMs $\{E_{(o)k}\}$ and
$\{F_{(o)\ell}\}$ generalizing the ones given in (\ref{7.9.1})'.
Then a bivariate Wigner measure may be defined (e.g. (\cite{dM2002}, \S7.9.3) according to
 \begin{equation}
 W_{(o)k'\ell'}(T):= \sum_{k\ell} \lambda^{-1}_{k'k} \mu^{-1}_{\ell'\ell} M_{(o)k\ell}(T),
 \label{7.9.2.1}
 \end{equation}
in which $\{\lambda^{-1}_{k'k}\}$ and $\{\mu^{-1}_ {\ell'\ell}\}$ are the inverses of
the nonideality matrices $\{\lambda_{kk'}\}$ and $\{\mu_{\ell\ell'}\}$, respectively.
Assuming the existence of these inverses, it can straightforwardly be proven that
\begin{equation}
\begin{array}{l}
\sum_{\ell'} W_{(o)k'\ell'}(T) = E_{(o)k'},\\
\sum_{k'} W_{(o)k'\ell'}(T) = F_{(o)\ell'}.
\end{array}
\label{7.9.1xx}
\end{equation}
Hence, if the nonideality matrices
$\{\lambda_{kk'}\}$ and $\{\mu_{\ell\ell'}\}$ are known, then
it is possible to find the \emph{ideal} (\emph{non}negative) probabilities $Tr_{(o)}\rho_{(o)}(t=0) E_{(o)k}$ and $Tr_{(o)}\rho_{(o)}(t=0) F_{(o)\ell}$
from the Wigner measure (\ref{7.9.2.1}) by calculating $Tr_{(o)} \rho_{(o)}(t=0) W_{(o)k'\ell'}(T)$
from the probability distribution $Tr_{(o)} \rho_{(o)}(t=0) M_{(o)k\ell}(T)$
(\emph{experimentally} obtained in a \emph{joint nonideal} measurement of two standard observables ($\{E_{(o)k}\}$ and
$\{F_{(o)\ell}\}$)).
In \cite{dM2002}, \S7.9.4 this possibility was seen to be extensible to `joint nonideal measurements of
\emph{more than two} standard observables', enabling `determination of the density operator $\rho_{(o)}(t=0)$
of the initial state of object (o)'
if a \emph{sufficient} number of PVMs (a so-called \emph{quorum} \cite{BandPark71,BandPark79}) is involved.

It is stressed here that `negativity of the quantities $Tr_{(o)} \rho_{(o)}(t=0) W_{(o)k'\ell'}(T)$ ' is
\emph{not} a reason to follow Diophantus
when around 300 p.C. calling a `\emph{negative} number' \emph{absurd} (e.g. https://nrich.maths.org/5961).
As a matter of fact, although having an \emph{epistemological} meaning (rather than the \emph{ontological} one
quantum mechanical probabilities $Tr_{(o)}\rho_{(o)}(t=0) M_{(o)k\ell}(T)$ are having since they are representing
`\emph{experimentally obtained} relative frequencies of measurement results'),
there is no reason to doubt applicability of negative Wigner measures within \emph{mathematical calculations} of the quantum mechanical theory.
Wigner measures may be especially useful elements of `computations performed by quantum computers', even if \emph{lacking}
a \emph{direct ontological} meaning.\\

{\bf{The polar/Schmidt decomposition and its `physical \emph{in}applicability'.}}

The \emph{positive} verdict found above with respect to `Wigner measures' will now be contrasted to
the following \emph{negative} verdict with
respect to the so-called `polar/Schmidt decomposition as applied to quantum measurement (e.g.~\cite{dM2002}, eq.~(1.59))'.
It has been proven by Schmidt \cite{Schmidt1906} that \emph{any} quantum mechanical \emph{two}-particle state
$|\psi_{12} \rangle = \sum_{ij} c_{ij} |\phi_{1i} \rangle  |\phi_{2j}\rangle$, --$\{|\phi_{1i} \rangle\}$ and
$\{|\phi_{2j}\rangle\}$ two \emph{arbitrary} `complete sets of mutually orthonormal vectors in \emph{different} Hilbert spaces
${\cal H}_k, k =1,2$ '--,
can be expressed according to $|\psi_{12} \rangle = \sum_i c_i |\phi^{(S)}_{1i} \rangle |\phi^{(S)}_{2i} \rangle$,
in which vectors $|\phi^{(S)}_{1i} \rangle$
are the eigenvectors of the density operator $\rho_1 = Tr_2 |\psi_{12} \rangle\langle \psi_{12}|$ (and, analogously, for
$|\phi^{(S)}_{2i} \rangle$).

In the present author's view (see also Ekert and Knight {\cite{EkertKnight95})
the `\emph{mutual dependence} of the sets $\{|\phi^{(S)}_{1i}\rangle\}$ and $\{|\phi^{(S)}_{2i}\rangle\}$
referred to above' is sufficient to bring in a `\emph{physical} verdict of \emph{in}applicability of the Schmidt decomposition'.
As a matter of fact, no `\emph{experimental} physicist' would accept that
a `\emph{purely mathematical possibility} of the Schmidt decomposition' might have a `physical \emph{counterpart}', since
it would require that there is a `\emph{deterministic correlation} (expressed by the Schmidt decomposition)
of \emph{pointer positions} (i.e. properties of the ancilla $(a)$) and \emph{measured properties of the (sub)microscopic object $(o)$)}'.
Such correlation would \emph{require} that, when changing
the `\emph{preparation procedure of the (sub)microscopic object}', the `\emph{measuring instrument would have to be changed too}',
thus requiring an \emph{un}physical \emph{correlation} between `preparation' and `measurement'.
For this reason \emph{within the domain of application of quantum mechanics} the `polar/Schmidt decomposition is
\emph{in}applicable to \emph{measurement}'.

\section{\bf{Discussion and conclusions}}
\label{sec3}

In this section my approach of `measurement within the \emph{(sub)microscopic} physical domain' is contrasted with
the so-called `quantum mechanical measurement problem'.
The idea, presented above, of restricting `application of the \emph{quantum mechanical theory}' to `measurement
on the \emph{(sub)microscopic physical domain}'
is compared to the `Copenhagen interpretation of quantum mechanics', as well as to `Bohr's ideas (which
are only \emph{partly} in agreement with the latter interpretation)'. In particular, influences from
\emph{philosophy} and \emph{mathematics} will be critically considered. In doing so I will restrict myself to measurements of which the POVMs $\{M_{(o)k}(T)\}$ are referring to `\emph{non}ideal measurements of PVMs'.\\

\subsection{\bf{Philosophical preamble}}
\label{subsec3.1}

This paper starts from a \emph{pragmatist} philosophy of physics.\footnote{For instance, Douglas McDermid \cite{McDermid}
(note, however, that I shall \emph{not} use the word `truth' if a `proposition is functioning satisfactorily',
but instead use `applicability'); see also e.g.~Suppe \cite{Sup77},\cite{Rescher77}.
The notion of `pragmatism/pragmatist' should be distinguished
from `pragmatics/pragmatic', the latter referring to the `phase of trial and error going with the inception of any physical theory'
(e.g. Stegm\"uller \cite{Stegmueller79}, Balzer, Sneed, Moulines \cite{BalSneMoul2000}, Kuipers \cite{Kuipers2001}; also
https://plato.stanford.edu/entries/physics-structuralism/).}
The main feature of the `\emph{pragmatist} philosophy of physics as entertained here' is the cutting of a Gordian knot by
`\emph{starting from a particular physical theory} (i.c. \emph{quantum mechanics})' and
`\emph{determining its domain of application}',
the `theory of everything' being deemed to be (at least presently) unattainable.
`Quantum mechanics' is assumed to have as its domain of application the \emph{(sub)microscopic} physical world
of atoms, electrons, photons, etc.,
\emph{encompassing} the `(sub)microscopic parts of the measuring instruments (probes/ancilla's)
necessary for obtaining knowledge on these objects'.

In the present paper `\emph{pragmatic} freedom' is applied when within quantum mechanics sticking to the
\emph{non-}relativistic Schr\"odinger equation (including (\ref{eq:2.8.10})), thus allowing to \emph{ignore} a certain necessity of
applying a `\emph{relativistic} counterpart of the latter equation'\footnote{For instance,
the Klein-Gordon equation.} necessary to eliminate a `controversy on the physical meaning of the Bell inequalities'
with respect to the question of `whether, or not, violation of the latter inequalities is caused by \emph{non}-local interactions'.
Instead, I shall abide with the by now growing acknowledgement that `violation of the Bell inequalities is a \emph{local} affair'
(e.g. de Muynck \cite{dM84,dM95,dM1986,dMFOP86,dM96}, Griffith \cite{Griff2020}),
as well as with the `feasibility of a distinction between
the notions of macro- and micro-locality/causality as developed in \cite{dM84}', only \emph{macro}locality being \emph{observable} within
the `domain of application of quantum mechanics' (see also Hegerfeldt et al.~\cite{HR}).

\subsection{\bf{Farewell to the `human observer'}}
\label{subsec3.2}

In the history of physics the human \emph{observer} has played an important role. In the present paper
this role is minimized as far as the `domain of application of \emph{quantum mechanics}' is concerned.
As a matter of fact, the `\emph{(sub)microscopic objects} quantum mechanics
is dealing with' are \emph{not} at all liable to be observed by the human observer. Within that domain \emph{measuring instruments}
are indispensable in order to enable human observation. In general an observer is just observing the \emph{positions of the pointers of his
measuring instruments},\footnote{Note that this also applies to e.g. thermodynamics and electrodynamics, where, respectively, thermometers,
Ammeters and Volt meters have to be used in order to obtain \emph{numerical} data.}
or, perhaps, is even just looking at a `graph produced by his printer after the measurement results have been processed by his computer'.
In particular is it deemed to be highly improbable that
the human observer is able to kill Schr\"odinger's cat by just looking at it.\footnote{Note that according to Schr\"odinger
``The first atomic decay would have poisoned it.'' (viz. by means of triggering a hammer
shattering a flask of hydrocyanic acid (\cite{Schroed35}, p.~157)). But this would just change the role of the human observer
into `observing \emph{another object}'.} In the present paper the role of the human observer is
thought to be taken over by the measuring instrument/probe (see \S\ref{subsubsec2.1.1} for an example of the way the `amplification
of the physical signal' --necessary for enabling human observation-- may be realized).

\subsection{\bf{Farewell to the `\emph{individual}-particle interpretation'}}
\label{subsec3.3}
Analogously to classical mechanics, the quantum mechanical wave function has often been interpreted as a description
of an \emph{individual} object, rather than as a description of an `\emph{ensemble} of (sub)microscopic objects'
(like e.g. a `beam of particles produced by a cyclotron', or `photons in a ray produced by a star').
However, it should be realized that at the basis of the `\emph{individual}-particle interpretation of
quantum mechanics' there is a \emph{misunderstanding}, to the effect that
the fundamental \emph{difference} is \emph{not} recognized between a `de Broglie wave (accompanying an \emph{individual} particle
(like a bow wave accompanying a ship))' and a `Schr\"odinger wave' (describing an \emph{ensemble}).

Today the `\emph{statistical} character of quantum mechanics' has been experimentally
corroborated,\footnote{for instance, https://www.hitachi.com/rd/research/materials/quantum/doubleslit/index.html, press: video clip 1,
Tonomura et al.~\cite{ToEnMaKaEz89}; also \cite{dM2002}, p.~202, Figure 4.4.}
\emph{in}dependently of whether the quantum mechanical state is described by a wave function/state vector or by a
density operator, or whether the measurement is described by a PVM or a POVM.
In the present paper the `\emph{ensemble} interpretation' is applied in its most general form.
Quantum mechanics is generally accepted by physicists as a \emph{statistical} theory, in which --analogously to
the general `\emph{im}possibility of predicting the digit an \emph{un}biased dice will show up after rolling on a table'--
within the domain of application of \emph{quantum mechanics} it is thought to be in general \emph{im}possible
to predict \emph{individual} measurement results.

In \emph{experimental} practice quantum mechanical probabilities are determined by \emph{repeating} an
`\emph{individual} execution of a measurement' a large number $N_{(exp)}$ (cf.~(\ref{eq:3.1.5.x})) of times,
and comparing the `probabilities/relative frequencies ${\bf p}_{(a)k}(T)$ of measurement
results corresponding to post-measurement pointer positions $k$ of ancilla $(a)$  (cf.~(\ref{eq:3.1.2}))' with the
experimentally obtained relative frequencies $\lim_{N_{(exp)} \rightarrow \infty} (N_{(exp)k}$/$N_{(exp)})$
(while taking `existence of this limit' as a
`\emph{necessary} requirement to be within the domain of application of the \emph{statistical} theory').
Note also that `$M_{(o)k}(T)$'s dependence on $T$ as defined by (\ref{7a.2.1.2})' is a \emph{symbolic} expression of
its `physical meaning as agreed upon in sections \ref{sec1} and \ref{sec2}'.

Quantum theory leaves open a complete answer to the question of whether quantum measurement is either
causal or noncausal, but is recognizing a `\emph{possibly in}complete causality' as implemented by
the conservation laws of energy and momentum as illustrated in \S\ref{subsubsec2.1.1}.

\subsection{\bf{Farewell to the `\emph{classical} paradigm'}}
\label{subsec3.4}

In the present paper the notion of `\emph{classical} paradigm' is implying that (like in most textbooks of \emph{classical} mechanics)
`\emph{also within quantum theoretical accounts measuring instruments are ignored}', and
`quantum mechanical measurement results (often represented by eigenvalues $a_{(o)i}$ of self-adjoint/Hermitian operators
$A_{(o)} = \sum_i a_{(o)i} |a_{(o)i}\rangle\langle a_{(o)i}|$)'
are treated as `\emph{properties of the (sub)microscopic object} $(o)$', rather than as
the `\emph{pointer positions of a measuring instrument}/ancilla $(a)$ they \emph{experimentally} are referring to'.
Analogously to \emph{classical} mechanics, most textbooks of \emph{quantum} mechanics are cast into the language of
`properties of the \emph{physical object under discussion}', while \emph{disregarding} the
`\emph{interaction of (sub)microscopic object $(o)$ and measuring instrument/probe} $(a)$'.
This is a source of misunderstandings, to be encountered in `\emph{realist} interpretations of both standard and
generalized versions of the quantum mechanical formalism'.

In the present paper the necessity and feasibility is discussed of `\emph{abandoning} the \emph{classical} paradigm'
analogously to the way we in \S\ref{subsubsec2.1.1} found this to be necessary in case of the Stern-Gerlach measurement.
The \emph{necessity} is stressed of \emph{within the (sub)microscopic physical domain}
replacing the `\emph{classical} paradigm' by a `\emph{modern} one', the latter paradigm taking into account
the \emph{interaction of a `(sub)microscopic physical object $(o)$' and `an equally (sub)microscopic part of a
measuring instrument/probe $(a)$'} (cf.~\S\ref{sec1}).

The present paper is discussing necessity, feasibility, and usefulness of `\emph{abandoning} the \emph{classical} paradigm'
by taking into account the `\emph{interaction} of the (sub)microscopic object $(o)$ with a measuring instrument/probe $(a)$'
in the way presented in \S\ref{sec1}.
In order to do so it is necessary to stress the \emph{different physical meanings}
probabilities ${\bf p}_{(o)k}$ and ${\bf p}_{(a)k}(T)$ (cf.~eq.~(\ref{eq:3.1.3.2.x})) are having notwithstanding their \emph{numerical equality}.
It is important to realize that it is \emph{not possible} to simply \emph{equate} a `\emph{measurement result obtained
by means of an interaction with a probe} $(a)$ ' to a `\emph{property of the object} $(o)$'.\footnote{As is done, for instance,
in a `length measurement by means of comparison with a measuring rod'.}
As a matter of fact, `what we may be able to see' is \emph{not} the `(sub)microscopic object
(with Stern-Gerlach being the spin of (an electron of) the atom)',
but it rather is a `property of the \emph{measuring instrument/probe's pointer}' (with Stern-Gerlach being physically
represented by a `spot caused by an atom's impact on a fluorescent screen').\footnote{The analogy
should be noted with measurements like `Kirchhoff's classical electric current measurement by means of an Ammeter'
as well as `measurement of temperature by means of a thermometer', which both might be seen as satisfying a \emph{modern} paradigm.}

In the \emph{empiricist} interpretation of quantum mechanics, as applied in \S\ref{subsubsec2.1.1} to Stern-Gerlach measurements,
the `atom's linear momentum $\vec{P}$ ' is taken as the \emph{pointer}
observable,\footnote{Note that in the SG measurement \emph{only} the atom's \emph{deflection} is experimentally determined;
the energy of its impact with the screen is left \emph{un}determined.}
and use is made of the `feature that $\vec{P}$ is \emph{not} a conserved quantity
but is \emph{changed} (be it (sub)microscopically) \emph{by the measurement interaction}'
in a way distinguishing --be it \emph{non}ideally
(cf.~\S\ref{subsubsec2.1.2})-- between (eigen)values of $\sigma_z$.
Amplification to macroscopic dimensions is realized by the `\emph{rectilinear} motion
into the direction the atom's momentum had obtained  \emph{after} it left the magnetic field',
thus allowing to \emph{distinguish} experimentally --be it \emph{non}ideally--
between \emph{different} eigenvalues of $\sigma_z$
by observing impacts of the `atoms of the ensemble' at \emph{different} positions of a fluorescent screen.

In the Stern-Gerlach measurement (cf.~\S\ref{subsubsec2.1.1}) phase~2 (defined in \S\ref{sec1}) was seen to be
the \emph{essential} part,
in which `information on the atom' is gained by letting it interact with a magnetic field,
thus changing its linear momentum in a way \emph{depending on the direction of the `spin of an electron of the atom'} (cf.~(\ref{7a.2.1.3})).
Amplification is realized in two phases (viz.~2 and 3), only
phase~2 (dealing with the `atom's interaction with the magnetic field') being discussed here,\footnote{Phase~3
(dealing with the atom's `interaction with a fluorescent screen') in experimental practice being dealt with \emph{semi-classically}.}
the atom's linear momentum being changed so as to cause the `impact position on the fluorescent screen' to depend on
`whether the \emph{initial} spin direction was $+$ or $-$'.
It is rather obvious that \emph{not the `measured object $(o)$ (i.e. the spin)'} is directly observed,
but rather `\emph{probe $(a)$'s position} when impacting on the screen'
(which position is determined by the \emph{direction} of the probe's linear momentum)
(cf.~\cite{MadM93}).\footnote{Note that the atom's kinetic energy must be sufficiently large in order that fluorescence be
visible. Note also that the \emph{quantum mechanical} description may allow a certain \emph{non}-ideality of
the measurement results (cf.~eqs (\ref{7a.2.1.1},\ref{7a.2.1.1xx})).}

Within `\emph{generalized} quantum mechanics as considered in \S\ref{sec2}'
it is possible to \emph{formally} maintain the `\emph{classical} paradigm' by \emph{restricting} attention to
probabilities ${\bf p}_{(o)k} = Tr_{(o)} (\rho_{(o)}(t=0)M_{(o)k}(T))$ defined in
eq.~(\ref{eq:3.1.3.2.x}), as was done --in agreement with a general custom of that
time-- in de Muynck \cite{dM2002},
an `\emph{empiricist approach}\footnote{Compare \cite{dM2002}'s title.}' being employed as a
`\emph{pragmatic} means to \emph{maintain} the \emph{classical paradigm}'.
POVMs were actually introduced in this latter way,\footnote{cf.~Davies and Lewis
\cite{DaviesLewis1970}, also \cite{Hel76} Chapt.~III.} and important `treatises
dealing with the properties of that new \emph{mathematical} entity' were written (e.g.~\cite{povm}),
thus \emph{seemingly} enabling the `\emph{classical} paradigm to be applied on the (larger) domain of POVMs'.
However, it should be realized that, whereas a measurement's goal is `\emph{experimental} determination of
\emph{pre}-measurement probabilities ${\bf p}_{(o)k}$ of properties of the \emph{(sub)microscopic object} $(o)$',
we, \emph{instead}, are \emph{experimentally} observing
`\emph{post}measurement probabilities ${\bf p}_{(a)k}(T)$ of \emph{pointer positions of a measuring instrument/probe $(a)$}'
(compare eq.~(\ref{eq:3.1.3.2.x})).

\emph{Within the `classical paradigm'} the \emph{physical} difference of POVMs and OVMs is usually left \emph{un}discussed.
For instance, notwithstanding his painstaking exploration of
`the basic structures of a physical theory', Ludwig's formidable work \cite{Ludwig67,Ludwig78,Ludwig83}
on the `Foundations of quantum mechanics' failed to `\emph{quantum mechanically} deal with
the \emph{interaction} of (sub)microscopic object and probe' (e.g.~\cite{Ludwig83}, Chapt.~XVII).
In most textbooks `quantum mechanical measurement results' are considered to refer to
`properties of the \emph{(sub)microscopic object} $(o)$', often even `\emph{without distinguishing}
between \emph{pre-} and \emph{post-}measurement ones'.

However, although $Tr_{(o)}(\rho_{(o)}(t=0) M_{(o)k}(T))\geq 0$  $\forall \rho_{(o)}(t=0)$ (as defined in (\S\ref{eq:3.1.1})),
this does \emph{not} imply that
$\{M_{(o)k}(T)\}$ would be a POVM  on \emph{all} states $\rho_{(o)}(t=T)= Tr_{(a)} \rho_{(oa)}(t=T)$.
In agreement with `Wigner's \emph{negative} probabilities (cf.~\S\ref{subsubsec2.2.2})' $p_{(o)k}$ may occasionally have
\emph{negative} values.
If the \emph{physical in}equality of $p_{(o)k}$ and ${\bf p}_{(a)k}(T)$ is \emph{not} taken into account, then the `change of the set of quantum mechanical observables from PVMs to POVMs $\{M_{(o)k}(T)\}$'
is liable to be considered as `just referring to a \emph{mathematical} generalization of the definition of
a quantum mechanical observable \emph{of the (sub)microscopic object $(o)$} (as is done in most textbooks of quantum mechanics)'
\emph{without} recognizing the role the
`interaction of object $(o)$ and ancilla/probe $(a)$' is playing in measurements of such objects.

This was duly remarked in a paper by Busch and Lahti \cite{BuLa96} (cf.~that paper's Abstract):
``The standard model of the quantum theory of measurement is based on an interaction
Hamiltonian in which the observable to be measured is multiplied by some
observable of a \emph{probe system} [My emphasis, WMdM]. This simple Ansatz has proved extremely fruitful in
the development of the foundations of quantum mechanics. While the ensuing type
of models has often been argued to be rather artificial, recent advances in quantum
optics have demonstrated their principal and practical feasibility.''
And: ``It is remarkable that for a long time the quantum measurement theory
and the theory of positive operator-valued measures developed quite independently
and largely without taking notice of each other. \ldots Fortunately the two
have since been brought together into a fruitful interaction.'' (l.c.~p.~891).

The aim of the present paper is to continue the above-mentioned development, addressed by Busch and Lahti, to take into account
the \emph{quantum mechanical interaction of (sub)microscopic object and measuring instrument}
(see also de Muynck (\cite{dM2002}, cf.~\S3.3.1, also \cite{NeoCop2007}).
It should be stressed, however, that, even though in the latter references it is realized that
`what we see is \emph{not} the (sub)microscopic object' but the `pointer of a measuring instrument',
within the \emph{empiricist} interpretation of quantum mechanics
the focus was largely on `properties of the \emph{(sub)microscopic object $(o)$}',
thus ignoring the importance of ${\bf p}_{(a)k}(T)$ by restricting all attention to ${\bf p}_{(o)k}$ (cf.~(\ref{eq:3.1.3.2.x})).

It seems to me that by ignoring the `dynamics of the \emph{measurement process} as represented by (\ref{eq:3.1.1})'
a crucial feature of `measurement within the quantum mechanical domain' is neglected, thus leaving room for
postulates --like von Neumann's projection and Schmidt's decomposition ones-- to fill up the gap left by \emph{ignoring} the
dynamics of the measurement process as a `process being governed by the \emph{interaction} of (sub)microscopic object and probe'.
As is seen from (\ref{eq:3.1.3.2.x}), POVM $\{M_{(o)k}(T)\}$ is \emph{depending on that interaction}.\footnote{Note that
in the present paper `\emph{abandonment} of the \emph{classical} paradigm
by taking into account the interaction of (sub)microscopic object $(o)$ and measuring instrument/probe $(a)$'
is symbolized by `\emph{explicit} reference in POVMs $\{M_{(o)k}(T)\}$ to \emph{measurement} time $T$ '.}

Note that this \emph{crucial} feature is \emph{not} essentially different from the possibility of `determining the \emph{pre}-measurement
temperature of a gas by letting it interact with a thermometer and observing the length change of that thermometer's mercury
column'.\footnote{This analogy was already noticed by von Neumann (\cite{vN32} Ch.~VI.1 (p.~419) (but, unfortunately, \emph{not}
continued on the next page); see also van Fraassen \cite{vF91}, p.~209/210.}
As a consequence it is possible to look upon probabilities ${\bf p}_{(a)k}(T)$ and ${\bf p}_{(o)k}$
--notwithstanding their \emph{numerical} equality-- in two \emph{different} ways.
Thus, pointers of measuring instruments being
--after a necessary
amplification\footnote{Compare the Stern-Gerlach example in \S\ref{subsubsec2.1.1}.}-- humanly observable,
it is reasonable to attribute to eq.~(\ref{eq:3.1.2}) an \emph{ontological} meaning
as `relative frequencies $Tr \rho_{(a)}(t=T) M_{(a)k}$ of \emph{pointer positions} (corresponding to POVM $\{M_{(a)k}\}$)
in the \emph{post}-measurement state $\rho_{(a)}(t=T)$ of the \emph{probe}'.

However, we should be aware that the above comparison of `quantum measurement' with `thermodynamic measurement'
is a dangerous one, \emph{unless} it is taken into account that thermodynamics is describing `what we see by \emph{applying
measuring instruments}'. Pointer positions of these instruments should \emph{not} be confounded with
`properties of the measured \emph{object}'. Thus, the hight of a thermometer's mercury column, brought into contact
with the `object the temperature of which it is meant to be measured', is changing \emph{only} if
before the measurement the temperatures of object and thermometer were \emph{different} from each other.
The `nonideality of the Stern-Gerlach measurement met in \S\ref{subsubsec2.1.2}'
has its \emph{classical} precursor.

If it is felt to be important that our physical theories are to be corroborated c.q. falsified by `measurement',
then we need to distinguish between `measuring instrument' and `object'.
Even if it is possible to cast `observations' in terms of `properties of the
(sub)microscopic object $(o)$' (as is done in (\ref{eq:3.1.3.2.x})), it should be realized that this does \emph{not} mean that `interaction
with the measuring instrument' would \emph{not} play a role.
Remembering that \emph{measurement} has the intention to obtain
`\emph{knowledge} on a (sub)microscopic object $(o)$ as it was \emph{prior} to measurement',
it is reasonable to attribute to $\{M_{(o)k}(T)\}$ an \emph{epistemological} meaning,
in the sense that `\emph{post}-measurement probability distributions of \emph{ontologically real}
pointer positions of a \emph{measuring instrument}'
may be representing `\emph{knowledge} on the \emph{pre}-measurement state $\rho_{(o)}(t=0)$ of the \emph{(sub)microscopic object} $(o)$'
(compare eq.~(\ref{eq:3.1.5.x})).

Theoretical physicists occasionally find it disappointing to restrict their attention to `just the phenomena' c.q. `just pointer
readings', and to have to forget about the `reality behind the phenomena' they --often justifiedly-- assume there to be.
For instance, John Bell: ``To restrict quantum mechanics to be exclusively about piddling laboratory operations is
to betray the great enterprise''.\footnote{John Bell, \emph{Against `measurement'} \protect{\cite{Bell90}}.}
However, in the present author's view the `\emph{pragmatist} methodology adhered to in the present paper' (cf.~\S\ref{subsec3.1})
is \emph{not at all restricting} to ``piddling laboratory operations'', \emph{nor} is it ``betraying the great enterprise''.
\emph{On the contrary}, by eqs~(\ref{eq:2.8.10}) through (\ref{eq:3.1.3.2.x}) ``laboratory operations'' are
\emph{in}cluded within the `theoretical description from which --due to the \emph{classical} paradigm--
they were largely \emph{ex}cluded! \emph{Un}observed interactions between two (sub)microscopic objects $(o)$ and $(o')$, say,
are liable to be described by eq.~(\ref{eq:2.8.10}) if $\rho_{(o)}(t=0)$ is replaced by $\rho_{(oo')}(t=0)$
(or $\rho_{(o)}(t=0)\rho_{o'}(t=0)$ in case the states of the two objects initially are \emph{un}correlated),
in which \emph{both} $(o)$ and $(o')$ may be \emph{arbitrary} (sub)microscopic objects, one of which being \emph{allowed
to be a probe}.\footnote{As already agreed upon in \S\ref{sec1}, within the mathematical formalism of quantum mechanics
the measuring instrument is restricted to its \emph{(sub)microscopic} part (i.e. the probe).}

`Measurement' is an important part of Bell's ``great enterprise''. Within the \emph{(sub)microscopic} domain the
`interaction between (sub)microscopic object and measuring instrument/probe' has to be taken into account.
In the present author's view `perpetuation of the \emph{classical} paradigm' is the main cause of the seemingly endless
discussion on the `quantum measurement problem', to be commented upon in the next
subsections.\footnote{For instance, even though in a recent paper \cite{WestWenn2019}
it is recognized that in the Stern-Gerlach experiment the magnetic field is \emph{in}homogeneous,
the description remains wanting due to sticking to the \emph{classical} paradigm.}
If the measuring instrument/probe
is given its legitimate position, then most bones of contention to be observed within quantum theory
are eliminated by the recognition that `measurement within the (sub)microscopic domain' is revealing `what we see
(i.e. the pointer positions of our measuring instruments)' rather than 'what there is (i.e.
the (sub)microscopic objects constituting the domain of application of quantum mechanics)'.

\subsection{\bf{Farewell to the `\emph{standard} formalism of quantum mechanics'?}}
\label{subsec3.5}

From \S\ref{sec1} --more specifically from
eq.~(\ref{eq:3.1.3.2.x})--
it is rather evident that the quantum mechanical formalism can \emph{not}
be restricted to \emph{standard} observables (corresponding to PVMs).
As a matter of fact, as seen in \S\ref{subsubsec2.1.2}, already one of the
`very first measurements to be described quantum mechanically (i.c. the Stern-Gerlach measurement)'
is needing a POVM.

This, however, does \emph{not} imply that `\emph{standard} quantum mechanics' should be completely abandoned.
If the nonideality matrix $\{\lambda_{kk'}\}$ (cf.~(\ref{7a.2.1.1x})) is invertible, it is possible to \emph{calculate} the \emph{ideal}
probabilities ${\bf q}_k=\sum_{k'} \lambda^{-1}_{k'k} {\bf p}_{(o)k'}$ from an experimentally obtained `\emph{non}-ideal
probability distribution $\{{\bf p}_{(o)k'}\}$' (compare \S\ref{subsubsec2.2.2}).
Therefore we have no reason to ignore the `abundant corroborations of
\emph{ideal} measurements of \emph{PVMs}' on which `restriction to \emph{standard} observables' is based,
even if --like in \emph{classical} mechanics-- usually the results are subjected to
(often considerable) `measurement errors'. \emph{Standard} measurements may remain involved,
be it that the `set of POVMs having $\lambda_{kk'}=\delta_{kk_0}$ for some $k_0$' is just a measure-$0$ \emph{sub}set of
the `set of all POVMs satisfying (\ref{7a.2.1.1x})'.

Evidently, if the nonideality matrix $\{\lambda_{kk'}\}$ is invertible,
there may exist \emph{non}ideal versions of `measuring an observable corresponding to a
particular PVM', yielding the \emph{same statistical information} as an ideal one.
However, it would be as unattractive to abolish `standard quantum mechanics' because of its
measure-$0$ status, as it would be unattractive to abolish the `\emph{classical} theory of \emph{point-like} objects
because \emph{point}-like objects are unphysical'. It might be advantageous, though,
to take a `theory of \emph{pointlike} objects' in an \emph{epistemological} --rather than an \emph{ontological}--
sense (see also \S\ref{subsec3.6}).

\subsection{\bf{Bohr versus Heisenberg, epistemology versus ontology}}
\label{subsec3.6}

Bohr was probably the first one to realize that `\emph{measurement}' must be a \emph{crucial} element of `\emph{any} physical theory
dealing with \emph{(sub)microscopic} physics'.
This certainly is a step into the direction of `\emph{abolishing} the \emph{classical}
paradigm (in \S\S\ref{subsec3.2} through \ref{subsec3.5} deemed to be necessary within the quantum mechanical domain)'. However,
in this author's view Bohr's step was too small. He only referred to a `\emph{macroscopic} measurement \emph{arrangement} as being a
boundary condition \emph{not} having any dynamics of its own', in particular `\emph{not possessing any pointer} enabling it to serve as a
measuring \emph{instrument}'.

A \emph{quantum mechanical} account of \emph{measurement}, like the one presented in \S\ref{sec1}, is not available in such an approach.
At most, a notion of \emph{contextuality} may have been acceptable to Bohr, to the effect that
`measurement results' are \emph{not} corresponding to `\emph{properties of a measuring instrument} (as is the case in
\S\ref{sec1})', but are thought to correspond to `\emph{properties of the (sub)microscopic object}, allegedly being
well-\emph{defined}\footnote{The issue of `unambiguous definition'
is raised by Bohr already in his Como lecture \cite{Bohr1928}. See also Bohr \cite{Bohr82}, p.~237-8.}\emph{only
within the context} of a \emph{particular} measurement arrangement'.
This latter choice seems to have been actually made by Bohr.

As Mermin \cite{Mermin2004,Mermin2014}
has convincingly argued, Bohr's statements are not always crystal clear, and may show signs of
the difficulty of coming to grips with a physical domain different from the well-known classical one.
But Bohr's insistence on a `\emph{classical} account of the \emph{measurement arrangement}' (to be distinguished from a
`\emph{measuring instrument}') is pretty evident, and does certainly \emph{not} point into the direction
of any `interaction of a \emph{(sub)microscopic} object with a \emph{(sub)microscopic} part of a
\emph{measuring instrument}/probe liable to be treated \emph{quantum mechanically} (being accepted in \S\ref{sec1})'.

Nevertheless, Bohr should be honoured for realizing that, within the
\emph{(sub)microscopic}
domain, \emph{measurement} is an essential element which cannot be ignored in the way it during a long time has been ignored within
\emph{theoretical} quantum physics.
Probably, due to Bohr's conviction --laid down in his `correspondence principle', implying
that \emph{measuring instruments must be macroscopic}, and, hence, should be described \emph{classically}--
a `\emph{quantum mechanical} description of the \emph{interaction of (sub)microscopic object and measuring instrument}
(as assumed in \S\ref{sec1})' would probably have been anathema with Bohr.\footnote{For a discussion of Bohr's `correspondence principle'
see e.g. \cite{dM2002}, \S4.3. It should be noted that
it might be preferable to attribute to \emph{Heisenberg} the \emph{weak} form of the `correspondence principle  (l.c.~\S4.3.1)'
I actually attributed to \emph{Bohr} (l.c.~\S4.3.3), that qualification better fitting \emph{Heisenberg}'s \emph{complete} reliance on the
\emph{quantum mechanical} theory he developed.}

Interesting in this respect is Bohr's clash with Heisenberg over the latter's idea that `the boundary/cut between
(sub)microscopic object and measuring instrument may be \emph{arbitrarily} shifted',\footnote{cf.~Archives for the History of
Quantum Physics, American Philosophical Society, Philadelphia, Bohr’s Scientific Correspondence, Bohr to Heisenberg, 15 September 1935,
``I have experienced difficulties by trying to understand more clearly the argumentation in your article.
For I am not quite sure that I fully understand the importance you attach to the freedom of shifting the cut between
the object and the measuring apparatus.''} and that,
hence, the \emph{whole} measurement might allow a \emph{quantum mechanical} description.
It seems to me that Bohr's criticism of Heisenberg --being aroused by the former's `\emph{correspondence principle}'--
was certainly correct, as \emph{much later} was recognized by Heisenberg himself (compare
``We can only talk about the processes that occur when, through the interaction of the particle with some other system,
such as a measuring instrument, the behaviour of the particle is to be disclosed. The conception of the objective
particles has thus evaporated, not in the fog of some new, obscure, or not yet understood reality concept, but into
the transparent clarity of a \emph{mathematics} [My emphasis, WMdM] that represents no longer
the behavior of the elementary particles but rather our \emph{knowledge} [My emphasis, WMdM] of this behavior.''
(Heisenberg \cite{Heis1958}, p.~100)).

Rather than a `\emph{mathematical} issue', Bohr stressed an \emph{epistemological} aspect,
in later publications by him often referred to as an `\emph{epistemological lesson}'\footnote{e.g. Bohr
(\cite{Bohr48}, p.~317, also \cite{Bohr1950}, pp.~53-4); see also the abundant references to `epistemological lesson'
in Folse \cite{Folse85}.} to be learnt from `measurement within the (sub)microscopic domain',
to the effect that `the measurement arrangement is \emph{defining which} (standard) observable is actually measured'
(in particular used in connection with his idea of `complementarity'). Even though, as seen in \S\ref{sec1}, Bohr's
idea of an `\emph{epistemological} connotation' was perfectly correct (even \emph{in}dependently of the `complementarity' issue),
but was it, due to \emph{in}sufficient development of that time's quantum theory (see also \S\ref{subsec3.7}),
too early to be able to see all features which nowadays are available.

Unfortunately, with Bohr his `epistemological lesson' boiled down to a vision of \emph{complementarity} as involving `\emph{correspondence},
\emph{mutual exclusiveness of measurement arrangements}, and \emph{incompatibility of quantum mechanical (standard) observables}'
(accepted as a basis of the `Copenhagen interpretation of quantum mechanics').
It should be stressed, however,
that there is an important difference between Bohr's interpretation and the Copenhagen one, to the effect that --contrary to Bohr's \emph{epistemological} intention-- within the \emph{Copenhagen} view
`complementarity' is interpreted --in a purely \emph{ontological} sense-- in terms of `mutual \emph{disturbance} of
\emph{measurement} results of \emph{in}compatible \emph{standard} observables of the \emph{(sub)microscopic object}
if \emph{simultaneously} measured', by Heisenberg \cite{Heis30} thought to be expressed by the `Heisenberg
inequality'\footnote{Note that, although usually being referred to as `Heisenberg inequality', in the general form
displayed in (\ref{eq:3.1.10}) they were derived by Kennard \cite{Kenn} for `\emph{canonically conjugated} standard observables',
and by Robertson \cite{Rob29} for `\emph{arbitrary} standard' ones  (also de Muynck \cite{dM2002}, \S1.7.1).}
\begin{equation}
\Delta A \Delta B \geq \frac{1}{2} | \langle\psi| [A,B]_-|\psi\rangle|,\;A \;{\rm and}\;B \;{\rm standard\; observables}.
\label{eq:3.1.10}
\end{equation}

However, both Bohr and Heisenberg overlooked the possibility --offered by the `introduction of a (sub)microscopic \emph{probe}
\emph{into the quantum mechanical theory}'--
of within the (sub)microscopic domain admitting a `\emph{quantum mechanical} description of the \emph{interaction}
of object $(o)$ and measuring instrument/probe $(a)$' (as done in \S\ref{sec1}).
Admittedly, Bohr as well as Heisenberg referred to such an interaction.
However, looking upon `measuring instruments' as `\emph{macroscopic} objects (to be described \emph{classically})'
neither of them dealt with that interaction as a `feature of a veritable \emph{quantum mechanical} measurement process'.

It had to wait until 1970 in order to see that the above idea (viz. that the Heisenberg inequality (\ref{eq:3.1.10}) would be
referring to an \emph{ontological} mutual disturbance of `properties
of the \emph{(sub)microscopic} object represented by \emph{in}compatible standard observables $A$ and $B$ of that object'
if measured simultaneously) was demonstrated to be \emph{false}.
As a matter of fact, it was made clear by Ballentine \cite{Bal70} that `the Heisenberg inequality does \emph{not} have
the \emph{ontological} meaning attributed by the Copenhagen interpretation to (\ref{eq:3.1.10})', since the quantities
$\Delta A$ and $\Delta B$ are \emph{not at all} referring to any `\emph{simultaneous} measurement of $A$ and $B$',
but are standard deviations `obtained in \emph{independently performed measurements} of either of these observables \emph{separately}'.

As already seen in \S\ref{sec1} (compare eq.~(\ref{eq:3.1.5.x})), this does \emph{not} imply that
`measurement within the (sub)quantum domain' would not have an \emph{ontological} meaning \emph{too}.
However, the latter is \emph{not} described by the `Heisenberg inequality (\ref{eq:3.1.10})'.
Heisenberg's physical intuition with respect to `\emph{ontological} mutual disturbance' was perfectly reliable
(as, by hindsight, might have been expected on the basis of our experience discussed in \S\ref{sec1}).
Based on careful studies of a number of `thought experiments'
Heisenberg was able to find his `\emph{uncertainty} relations', which he, however --in agreement with
Bohr's `\emph{epistemological} lesson'-- in \emph{later} publications formulated
in an \emph{epistemological} sense.\footnote{For instance, in Heisenberg \cite{Heis30}, \S II.2
he is referring to ``\emph{knowledge} [My emphasis, WMdM] of a particle's momentum by an apparatus determining its
position'' (``Impuls\emph{kenntnis} [My emphasis, WMdM] durch einen Apparat zur Ortsmessung'' in the original German edition).}
His analysis (Heisenberg \cite{Heis30}) being hampered by the `\emph{classical} paradigm', and, hence,
\emph{no} `quantum mechanical description of the \emph{measuring instrument}' being contemplated by him,
Heisenberg was \emph{un}able to \emph{mathematically} account for the `\emph{interaction of (sub)microscopic object and
measuring instrument}' he took into account in a \emph{verbal} sense.

The quantum mechanical theory of measurement (as discussed in \S\ref{sec1}) being able to deal with
the `interaction of (sub)microscopic object $(o)$ and probe $(a)$', is allowing to describe `the \emph{(sub)microscopic} part of the
measurement process'. Nowadays having available a wealth of \emph{experimental} material
corroborating this theory (e.g. de Muynck \cite{dM2002}), we are able to take advantage of the
`\emph{extension of the quantum mechanical formalism by encompassing the measuring instrument/probe $(a)$}',
thus obtaining the generalization of the concept of `quantum mechanical observable of (sub)microscopic object $(o)$'
to, for instance, the `POVM $\{M_{(o)k\ell}(T)\}$ satisfying relations (\ref{7.9.1xxx})'.
But, as a result of his acceptance of the `classical paradigm', Heisenberg\footnote{Followed by many ``Copenhageners''.} was
\emph{un}able to interpret \emph{correctly} the inequalities named after him (cf.~the `Addition in proof' in Heisenberg \cite{Heis27},
in which the latter is confessing that ``Bohr has brought
to my attention that I have overlooked essential points in the course of several discussions in this paper.'').

Summarizing, we may conclude that the `tension often felt to exist within the interpretation of quantum mechanics'
between Bohr's \emph{epistemological} and Heisenberg's ((early) Copenhagen) \emph{ontological} `views on the physical meaning of quantum mechanics'
can be resolved by realizing that both of them \emph{partly} ignored the `crucial role of the measuring instrument',
Bohr \emph{not} acknowledging \emph{any ontological} influence of the \emph{measurement arrangement} on the (sub)microscopic object at all,
whereas Heisenberg allowed for such an influence, be it \emph{only a disturbing one} (as expressed by the
`Heisenberg inequality (\ref{eq:3.1.10})'). However, in agreement with
the `\emph{classical} paradigm' (which was abandoned in \S\ref{subsec3.4}) neither of them
accounted for the `\emph{influence of the (sub)microscopic object $(o)$ on the measuring instrument/probe} $(a)$', the latter nowadays
more and more being felt to be an important issue \emph{also} within the \emph{quantum mechanical} domain.

The `generalized Martens inequality (\ref{2.2.1.3})' can be seen as a generalization of the Heisenberg inequality (\ref{eq:3.1.10}),
taking into account \emph{not only} the `\emph{interaction of (sub)microscopic object $(o)$ and ancilla $(a)$}', \emph{but also}
the influence of the `\emph{preparation of the initial state of object} $(o)$'.

\subsection{\bf{von Neumann}}
\label{subsec3.7}

Let me start by agreeing that von Neumann was a brilliant mathematician/logician. For a whole generation of physicists
his book \cite{vN32} has been a standard introduction to the \emph{mathematical} intricacies of quantum mechanics.
However, not all of von Neumann's ideas have survived the development of quantum physics (both
experimental and theoretical) in the 90 years after the publication of his book. In particular his well-known `projection
postulate' is (also in the present paper) a bone of contention (e.g. de Muynck \cite{dM2002}).

In order to understand von Neumann's ideas with respect to `quantum mechanical measurement' it is necessary to take
into account that in his book he is drawing a fundamental distinction between `\emph{free evolution} of a (sub)microscopic object'
and `evolution of that object when \emph{subjected to measurement}' (cf.~\cite{vN32}, p.~417-8).
Thus, \emph{free}\footnote{`free' to be interpreted as `\emph{not} interacting with a
\emph{measuring instrument}'.} evolution of an object ($o$)'s quantum state $|\Phi_{(o)}(t)\rangle$
is assumed to be described by the Schr\"odinger equation
\begin{equation}
i\hbar\; d|\Phi_{(o)}(t)\rangle/dt = H_{(o)} |\Phi_{(o)}(t)\rangle,
\label{3.6.1}
\end{equation}
having as solutions
\begin{equation}
|\Phi_{(o)}(t)\rangle = U_{(o)}(t)|\Phi_{(o)}(t=0)\rangle,\; U_{(o)}(t)= e^{-iH_{(o)}t/\hbar}.
\label{3.6.2}
\end{equation}
Von Neumann also was aware of the possibility of using \emph{density operators} $\rho_{(o)}(t)$ (cf.~\cite{vN32}, p.~351) next to
\emph{state vectors} $|\Phi_{(o)}(t)\rangle$.

But it should be stressed that, with the exception of Chapt.~VI, von Neumann's book does \emph{not} transcend the
`\emph{classical} paradigm' (cf.~\S\ref{subsec3.4}) in the way that paradigm is transcended by eq.~(\ref{eq:2.8.10}) of the present paper.
As von Neumann is putting it in \cite{vN32},~p.~420: ``Now quantum mechanics describes the events which occur
in the observed portions of the world, so long as they do \emph{not interact with the observing portion} [my emphasis, WMdM],
with the aid of process {\bf 2.} (V.1.) [i.e. the Schr\"odinger equation, WMdM],
but as soon as such an interaction occurs, i.e. a measurement, it \emph{requires} [my emphasis, WMdM] the application of process {\bf 1.} (V.1.)
[i.e. von Neumann \emph{projection} (\cite{vN32}~p.~351), WMdM]''.
Von Neumann is drawing a fundamental difference between these two (what he calls) ``interventions'', to the effect that
according to him process {\bf 2.} is \emph{deterministic/causal}, whereas process {\bf 1.} is \emph{arbitrary/statistical}
(see also l.c.~p.~357).

Only in the very last section (\cite{vN32}, V.1.3) a `DISCUSSION OF THE MEASUREMENT PROCESS' is
contemplated in which an `\emph{interaction of object and measuring instrument}' is described \emph{quantum mechanically}.
Unfortunately, as it seems to me, von Neumann's discussion is not very illuminating because
--probably since he was following Bohr (cf.~\S\ref{subsec3.6}) while dealing with \emph{measuring instruments}--
he does \emph{not} consider any evolution equation referring to the `interaction of object $(o)$
and ancilla $(a)$ (as is dealt with by eq.~(\ref{eq:2.8.10}))'.
Instead, in the first paragraph of that section he is writing
that ``we shall make use of the results of V.1.2. to \emph{exclude} [my emphasis, WMdM] a possible explanation
often proposed for the statistical character of process {\bf 1.} (V.1.)''.

Taking into account that in \S V.I.2 von Neumann was discussing the `Schmidt decomposition' (displayed on pages 434 and 435),
it is my guess that he here is referring to this latter decomposition, in \S VI.3 demonstrating that a `result
mathematically equal to the Schmidt decomposition (cf. the last equation on p.~440)' can be \emph{derived as a consequence of his
projection postulate}. It should be stressed, however, that von Neumann's projection postulate is far from universally accepted.
Thus, in de Muynck (\cite{dM2002}, p.~23) it is concluded that
``Therefore, a conclusion of this book will be that the projection postulate, at least in the form
presented above, must be relegated to the realm of quantum mechanical folklore.'' Anyway, it is meeting the same
objection as the one concluded to in \S\ref{subsubsec2.2.2}, viz. that von Neumann's projection postulate is
\emph{ignoring} the `\emph{experimental} necessity of physical \emph{in}dependence of \emph{preparation} and \emph{measurement} processes'
(`preparation of object $(o)$' being restricted to phase~1 (cf.~\S\ref{sec1})).

It is remarkable that in \cite{vN32}, p.~421 and p.~439 von Neumann was close to the present paper's account,
when distinguishing between I (the system actually observed), II (the measuring instrument), and III (the observer,
who ``remains \emph{outside} [my emphasis, WMdM] of the calculations'').
But, while on p.~420 referring to Bohr's ``principle of the psycho-physical parallelism'' (cf.~Bohr \cite{Bohr1929}),
von Neumann ``\emph{required} [my emphasis, WMdM] application of Process (1.) as soon as such an interaction [i.e. a \emph{measurement} interaction, WMdM] occurs''.
He evidently was aware that the quantum mechanical description is a \emph{statistical} one, and that the human observer is free
to select \emph{sub}ensembles. He might even have been aware that such a selection may be carried out on the basis of
observations of `pointer positions of a measuring instrument', although this is improbable since Bohr's measurement arrangements
did not have pointers.

However, perhaps von Neumann was too much a mathematician (rather than a physicist)
to be aware that Schmidt's decomposition formula
is too restrictive to be useful for yielding an account dealing with the `\emph{preparation} of \emph{sub-ensembles}
conditional on individual measurement
results $k$ (cf.~eq.~(\ref{eq:3.1.5.x})) referring to pointer positions of a measuring instrument's probe',
`probabilities of which' in general \emph{not} being given by expectation values of the PVMs of the Schmidt decomposition,
 but by POVMs, an issue \emph{not} being raised \emph{without} the general
\emph{quantum mechanical} account of the `\emph{interaction} of (sub)microscopic object $(o)$ and measuring instrument/probe $(a)$
presented in \S\ref{sec1}'.\footnote{See also de Muynck \cite{dM2000},\cite{NeoCop2004}.}
As a matter of fact, had von Neumann in the very last sentence of his book not prematurely finished his discussion by
`leaving further discussion to the reader (\cite{vN32} p.~436)', then perhaps the present quest
--only \emph{partly} being presented in \cite{dM2002} but continued in the present paper--  would \emph{not} have been necessary.

The present paper is \emph{not} challenging the conclusions drawn by von Neumann and Bohr. The issues they raised are certainly
important. Yet there are reasons to doubt the \emph{universal} validity of their conclusions.
Nowadays experiment as well as theory is increasingly
able to find deviations from the schemes of `Bohr's correspondence principle', as well as from `von Neumann's projection postulate'.

\bibliographystyle{unsrt}

\end{document}